\documentclass{aa}  
\usepackage{graphicx}	
\usepackage{amsmath}	
\usepackage{amssymb}	
\usepackage{bm}	        
\usepackage{csvsimple}
\usepackage{ragged2e}
\usepackage{multirow}
\usepackage{siunitx} 
\usepackage[utf8]{inputenc}
\usepackage{dirtytalk}
\usepackage{xparse}
\usepackage{enumitem} 
\usepackage{array}
\usepackage{CJKutf8}
\usepackage[flushleft]{threeparttable}
\usepackage{txfonts}
\usepackage[colorlinks, citecolor=blue, linkcolor=blue]{hyperref}
\usepackage{graphicx}
\usepackage{txfonts}
\usepackage{svg}
\usepackage{orcidlink}
\newcommand{\orcid}[1]{\orcidlink{#1}}

\begin{document}

   \title{Emergence of high-mass stars in complex fiber networks (EMERGE)}

   \subtitle{VII. Multi-Tracer Column Density Synthesis}

\titlerunning{EMERGE (VII). Multi-Tracer Column Density Synthesis}

   \author{A. Hacar
          \inst{1}\orcid{0000-0001-5397-6961}
          \and 
          A. Socci\inst{1}\orcid{0009-0008-0541-572X}
          \and
          M. Tafalla\inst{2}\orcid{0000-0002-2569-1253}
          \and
          F. Bonanomi\inst{1}\orcid{0000-0002-7154-9629}
          }

   \institute{Department of Astrophysics, University of Vienna,
              T\"urkenschanzstrasse 17, A-1180 Vienna\\
              \email{alvaro.hacar@univie.ac.at}
         \and
             Observatorio Astron\'omico Nacional (IGN), Alfonso XII 3, 28014 Madrid, Spain
             }

   \date{Received XXX ; accepted XXX}

 
  \abstract
   {Obtaining accurate estimates of the gas column density $\mathrm{N(H_2)}$ is crucial to determine the physical conditions of the molecular gas in the Interstellar Medium. Standard dust extinction, far-IR emission, (sub-)millimeter continuum, and/or molecular lines observations regularly produce high-fidelity column density maps at low spatial resolution. However, sensitivity limitations hamper the use of these techniques at the higher interferometric resolutions.}
   {This paper introduces the "Multi-Tracer Column Density Synthesis" (MT-CDS) technique, a new empirical approach to derive high dynamic-range column density maps with interferometers such as ALMA.}
   {We have compared the integrated intensity maps of two density selective tracers, namely HNC (1-0) and $\mathrm{N_2H^+}$ (1-0), obtained with IRAM-30m with ancillary {\it Herschel} column density estimates in seven star-forming regions in Orion and derived individual temperature-corrected conversion factors for each of these tracers.
   We applied these conversion factors obtained at low-resolution (30" or $\sim12500$~au) into similar high sensitivity ALMA observations to derive independent column density maps describing both the low $N_\mathrm{HNC}\mathrm{(H_2)}$ and high $N_\mathrm{N_2H^+}\mathrm{(H_2)}$ column density components at high spatial resolution (4.5" or $\sim2000$~au).
   We also combined these independent maps into estimates for the total gas column density gas $N_\mathrm{{tot}}\mathrm{(H_2)}$ sampling more than two orders of magnitude in column density between $\sim 10^{21.5}-10^{23.5}$~cm$^{-2}$.
   A comparison with independent ArTeMiS column density maps at intermediate resolution (8" or $\sim3200$~au)
   shows typical deviations of $\lesssim$0.2~dex from our MT-CDS gas column density estimates.
   }
   {We obtained individual column density maps in five of our targets at $\sim$2000~au resolution in which ALMA observations were available, that is a factor of $\sim$5 better than originally calibrated.
   These results demonstrate the potential of this MT-CDS technique to retrieve accurate, high dynamic range $N\mathrm{(H_2)}$ total gas column density estimates at interferometric resolution combining multiple molecular line tracers. 
   }
   {The high resolution total column density maps obtained using this MT-CDS method reveal the hierarchical and multi-scale nature of the ISM with unprecedented detail. Our results highlight the need of high-resolution observations when interpreting the internal structure of molecular clouds.}

   \keywords{ISM: clouds -- ISM: kinematics and dynamics -- ISM: structure -- stars: formation -- ISM: molecules
               }

   \maketitle
%

\section{Introduction}\label{sec:intro}

Deriving accurate measurements of the total gas column density $N(\mathrm{H}_2)$ is of fundamental importance to characterize the physical properties of molecular clouds. Estimates of $N(\mathrm{H}_2)$ enable calculations of the total gas mass of molecular clouds as well as their gravitational stability. Column density maps facilitate the identification of the cloud substructure and organization into filaments and cores. Moreover, the mean and distribution of the total gas column density $N(\mathrm{H}_2)$ and the gas surface density of molecular clouds are closely related to the processes that govern their evolution and star formation efficiency \citep{Lada2010,Kainulainen2014_rhoPDF}. Consequently, obtaining high-fidelity, high dynamic range $N(\mathrm{H}_2)$ maps of molecular clouds near and far has become an observational priority since the first detections of these objects.

The total gas column density $N(\mathrm{H}_2)$ in molecular clouds can be inferred from dust extinction and emission, given basic assumptions about the dust cross section, reddening law, and dust-to-gas ratio (i.e. $N(\mathrm{H}_2)\propto A_V$).
Large-scale optical and near-IR (NIR) extinction surveys produced the first maps of the total gas column density in nearby molecular clouds down to $N(\mathrm{H}_2)\sim10^{21}\ \mathrm{cm}^{-2}$, at the expense of a coarse $\sim30$–$60''$ resolution \citep{Lada1999,Lombardi2001,Dobashi2005,Lombardi2006,Lombardi2009,Lombardi2011,Kainulainen2009,Meingast2018,Zhang2022_PROMISE,Zhang2025_PROMISE}.
Extinction observations in the mid-IR (MIR) allow to improve this resolution down to 2-5" in some favorable targets \citep{Kainulainen2013,Juvela2023}.

Estimates of $N(\mathrm{H}_2)$ can be also obtained from the observation of the dust emission $F_\nu$ in molecular clouds. Specifically, $N(\mathrm{H}_2)\propto F_{\nu}\propto B_\nu(T_\mathrm{dust})\cdot \tau_{dust,\nu}$ in the optically thin regime, where $T_\mathrm{dust}$ is the dust temperature and 
$\tau_{dust,\nu}\propto \kappa_0\cdot\left(\frac{\nu}{\nu_0}\right)^{\beta}$ is the optical depth parameterized by the
dust emission coefficient $\kappa_0$ and dust emissivity index $\beta$.
Ground-based single-dish maps of the dust emission at (sub)millimeter wavelengths can achieve resolutions of $\sim8$–$30''$, revealing internal cloud structure above $N(\mathrm{H}_2)\sim10^{22}\ \mathrm{cm}^{-2}$ but with limited sensitivity to large-scale emission due to spatial filtering \citep{Motte1998,Young2006_Bolocam,Schnee2007,Motte2007_MAMBO,Nutter2008,Kauffmann2008,Schuller2009,Hacar2013,Kirk2016,Kirk2018}.
Space telescopes such as Herschel and Planck have greatly advanced these results in angular scale, resolution, and sensitivity via multi-band far-IR (FIR) observations. Current studies routinely provide large dynamic range column density maps spanning $N(\mathrm{H}_2)\sim(1$–$100)\times10^{21}\ \mathrm{cm}^{-2}$ at cloud scales with $\sim15$–$30''$ resolution in nearby star-forming regions \citep{Andre2010,Palmeirim2013,Lombardi2014,Konyves2015,Lada2017,Hasenberger2018} and Galactic Plane surveys \citep{Molinari2010,Traficante2011,Molinari2016,Motte2010,Marsh2015_PPMAP}.

Radio interferometers such as ALMA provide dust continuum measurements at 0.5-5" resolution \citep[e.g.,][]{Sanhueza2019,Motte2021,Coletta2025_ALMAIMF}. However, the lack of short-spacing information in most of these observations critically hampers the detection of the extended continuum emission in resolved molecular clouds \citep{Ossenkopf2016,BonanomiPaperII}. While data combination with single-dish bolometric observations is feasible \citep{Ginsburg2026_CMZ,Xu2026}, sensitivity limitations usually restrict the dynamic range of most column density maps obtained by interferometric observations. 

Equivalent column density measurements can be obtained from molecular line observations. A primary advantage in this context is that selective molecular tracers and transitions can probe distinct gas density regimes in molecular clouds. Compared to the relatively homogeneous treatment of dust, however, using molecular lines for these purposes is intrinsically more complex. First, the observed line main beam brightness $T_{\mathrm{mb,\nu}}$ at a given frequency $\nu$ depends on the excitation temperature $T_{\mathrm{ex}}$, the background temperature $T_{\mathrm{bg}}$, and the optical depth $\tau_{\nu}$ of the chosen transition, following the radiative transfer equation (RTE) integrated over the line-of-sight, i.e. $T_{\mathrm{mb,\nu}}=\int f(T_{ex},T_{bg},\tau_{\nu})\cdot ds$.
Second, the observed properties are modulated by the local physical conditions, primarily gas density and kinetic temperature, since $T_{\mathrm{ex}}=f(T_K,n)$. Simplified treatments usually adopt Local Thermodynamical Equilibrium (LTE) conditions \citep[i.e. $T_{\mathrm{ex}}=T_K$; ][]{Frerking1982,Hacar2013}, although this assumption is likely not satisfied in most cases \citep{Shirley2015}. 
Third, the conversion between the specific molecular column density $N(\mathrm{A})=\int T_{mb,\nu}(T_{ex},\tau_\nu)\cdot d\nu $ and the total gas column density depends on the molecule’s relative abundance $X(\mathrm{A}/\mathrm{H}_2)$, so $N(\mathrm{H}_2)=N(\mathrm{A})\cdot X(\mathrm{A}/\mathrm{H}_2)$, where $X(\mathrm{A}/\mathrm{H}_2)$ is often poorly constrained. Resolving these uncertainties requires multiple line transitions (e.g., J-ladders and/or isotopologues) together with chemical and/or radiative transfer models, which limits the applicability of these methods to specific targets \citep{Teng2020} or dedicated surveys \citep{Shimajiri2014, Nishimura2015}.

Observational estimates of the total gas column density can be obtained using the so-called X-factors $X_i$, that is the empirical conversion factor between the observed integrated intensity $I=\int T_{mb,\nu}\cdot d\nu$ and the total gas column density such as $N(\mathrm{H}_2)=X_i\cdot I_i$. The X-factor encapsulates the above excitation, opacity, and abundance conditions of a given transition into a single average conversion factor over large physical scales. 
The use of $X_\mathrm{CO}$ and $X_\mathrm{HCN}$ conversion X-factors has become particularly popular in extragalactic studies at kpc-scales \citep[e.g.,][]{Kennicutt1998,GaoSolomon2004}. 
However, the known dependence of these X-factors with metallicity, density, and temperature \citep{Pety2017,Hacar2020,Tafalla2023,Dame2023} complicates their use to obtain high-precision column density estimates at parsec scales \citep[see ][for a full discussion about $X_\mathrm{CO}$]{Bolatto2013}.

The use of low-noise heterodyne receivers and advanced data reduction techniques produces high-dynamic range molecular emission maps at both single-dish \citep[5-30";][]{Pety2017,Kauffmann2017} and interferometric \citep[$\sim$1-5";][]{Hacar2018,Barnes2021,Beuther2022_CASCADE} resolution. On the other hand, the selective nature of most molecular tracers appears as main drawback of line observations compared to dust continuum maps. The combination of excitation, opacity, and chemical effects make specific molecular species and transitions sensitive only to a limited range in column (and volume) density. Among other classical tracers, low-J transitions of molecules such as C$^{18}$O, HCN, and HNC are known to be sensitive to low column density material between $N\mathrm{(H_2)}\sim [1,50]\times10^{21}$~cm$^{-2}$, while density selective species such as N$_2$H$^+$ and NH$_3$ only appear at $N\mathrm{(H_2)}> 10\times10^{21}$~cm$^{-2}$ \citep{Pety2017,Tafalla2021,Tafalla2023}.
To overcome these issues, novel approaches include the combination of multiple line diagnostics and transitions with non-LTE radiative transfer models \citep{Segal2024}, and inversion techniques \citep{Palud2025}. A promising new avenue includes the use of machine learning methods in multi-line surveys to retrieve the full column density structure of molecular clouds at high resolution \citep{Gratier2021,Shimajiri2023_H2}. Computationally and observationally expensive, these methods have been applied to few dedicated studies. 

This paper introduces the ``Multi-Tracer Column Density Synthesis'' (MT-CDS), a new observational technique
that explores the use of empirically calibrated temperature-corrected molecular conversion factors to obtain high-resolution column density maps at interferometric resolutions. 
We calibrate low-resolution (30") HNC~(1-0) and N$_2$H$^+$~(1-0) emission maps obtained at the IRAM-30m telescope against ancillary Herschel column density maps in seven star-forming regions in Orion sampled by the EMERGE Survey (Sect.~\ref{sec:EMERGE}). We derive unique temperature-corrected X-factors for HNC and N$_2$H$^+$, respectively, describing the emission of both low and high column density gas components in all our targets (Sect.~\ref{sec:Xfactors}).
We apply these X-factors into ALMA molecular maps of these same tracers to construct equivalent and independent column density maps of each of these gas components at high spatial resolution (4.5"). We also combine these maps to obtain a total (low plus high) column density map at similar resolution that closely reproduces previous dust total column density maps proving the validity of this new MT-CDS technique ( Sects.~\ref{sec:Xfactors_highres}-\ref{sec:accuracy}).
We produce high dynamic range and individual low, high, and total column density maps in the EMERGE five targets when both HNC and N$_2$H$^+$ ALMA maps were available (Sect.~\ref{sec:NH2colmaps}). Our results reveal the fine gas organization at $\sim 2000$~au scales (4.5" at the distance of Orion) only observable at interferometric resolution (Sect.~\ref{sec:tomography}). 


\section{Observations: EMERGE Early ALMA Survey}\label{sec:EMERGE}

\begin{figure*}[ht!]
\centering
\includegraphics[width=0.95\textwidth]{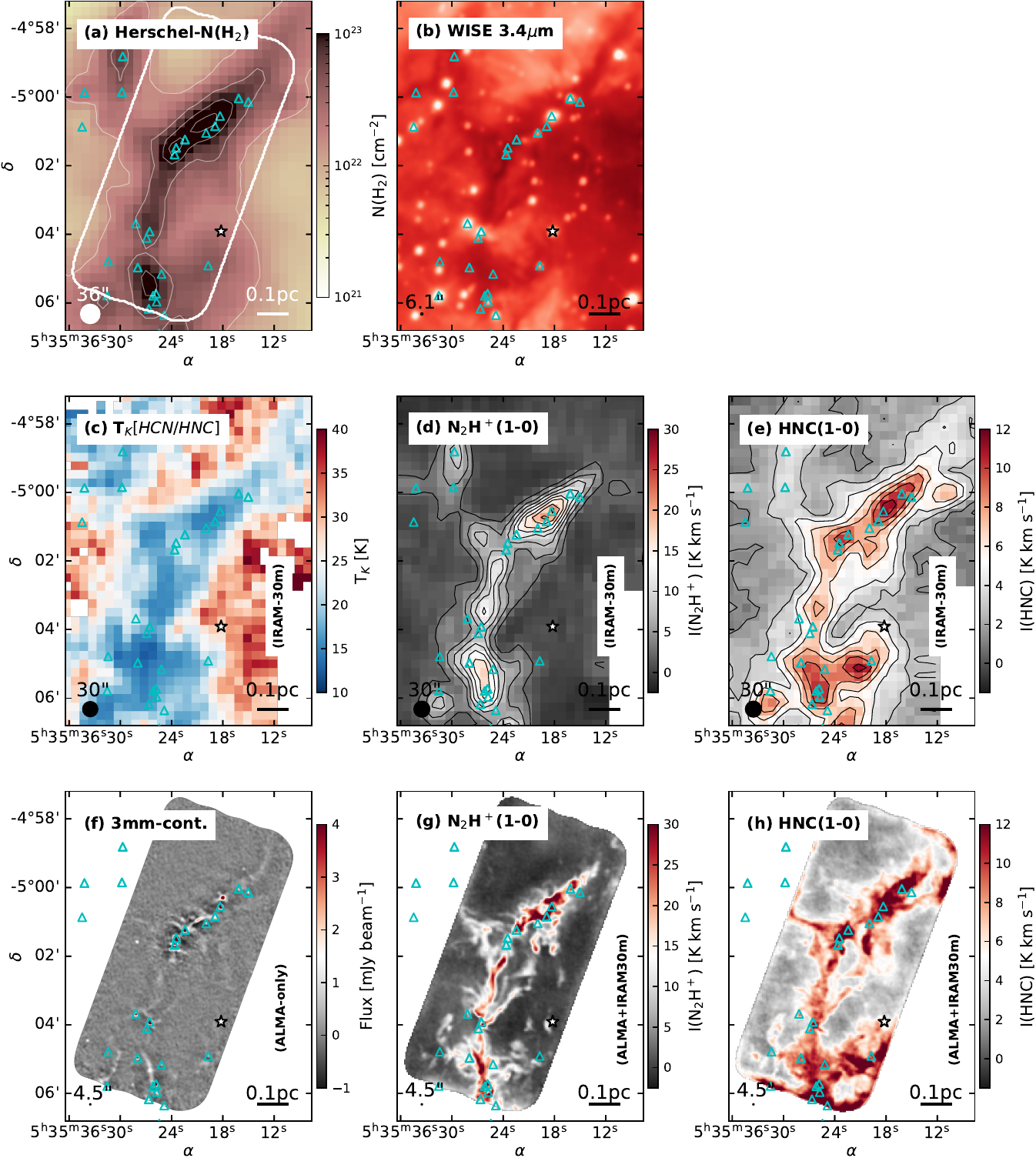}
      \caption{EMERGE Early ALMA survey observations in OMC-3. From left to right and from top to bottom we plot 
      {\bf (a)} Herschel, total gas column density, $N\mathrm{(H_2)}$ (36"); 
      {\bf (b)} {\it WISE} 3.4~$\mu$m emission (6.1");
      {\bf (c)} gas kinetic temperature, $T_\mathrm{K}$[HCN/HNC],
      {\bf (d)} N$_2$H$^+$ (1-0), and {\bf (e)} HNC (1-0) obtained with IRAM-30m observations (30");
      {\bf (f)} 3mm continuum emission (ALMA-12m only), plus
      {\bf (g)} N$_2$H$^+$ (1-0) and {\bf (h)} HNC (1-0) ALMA+IRAM-30m MACF high-resolution (4.5") maps.
      The positions of young protostars (cyan triangles) \citep[][]{Megeath2012} and OB-stars (white stars) are indicated in all panels. The corresponding beam size and scale bar are indicated in the lower corner of each panel. For reference, panel a includes contours at $N\mathrm{(H_2)}$ = [15, 50, 100, 150, 200] $\times 10^{21}$~cm$^{-2}$ and the footprint of our ALMA observations. Reference contours in the N$_2$H$^+$ (panel d) and HNC (panel e) IRAM-30m maps are equally spaced every 2.5 and 1.5~K~km~s$^{-1}$, respectively.
      Similar plots for all the other regions included in this study can be found in \citep{Hacar2024}.
      }
\label{fig:EMERGE_maps}
\end{figure*}

We make use of the first observations of the EMERGE project\footnote{EMERGE Project website: \url{https://emerge.univie.ac.at/}}.
In its EMERGE Early ALMA Survey \citep[][; hereafter Paper I]{Hacar2024}, this project targeted seven star-forming regions in Orion (OMC-1/2/3/4 South, LDN~1641N, NGC~2023, and Flame~Nebula) all mapped at high spectral resolution ($\delta v\le 0.25$~km~s$^{-1}$) by both ALMA-12m inteferometric (4.5") and IRAM-30m single-dish (30") observations in N$_2$H$^+$~(1–0), HNC~(1–0), and HCN~(1-0) (IRAM-30m-only) lines alongside 3~mm-continuum (ALMA only).
These ALMA-12m and IRAM-30m data were merged to produce high-sensitivity, high-dynamic range molecular maps of the above three molecular species at high spatial resolution (4.5" or $\sim2000$~au) using the Model Assisted Cleaning and Feathering (MACF) data combination technique (hereafter ALMA+IRAM-30m MACF; see Paper I for a full description).
These observations are complemented by low spatial resolution (30") gas kinetic temperature maps ($T_K$(HCN/HNC)) obtained from the observed line ratio between HCN~(1-0) and HNC~(1-0) transitions \citep{Hacar2020}, as well as ancillary {\it Herschel} column density maps obtained in all these regions at 36" resolution \citep{Lombardi2014}. 

We display a representative sample of these first EMERGE observations for the analysis of OMC-3 in Fig.~\ref{fig:EMERGE_maps}. The mapping strategy (see white contour in panel a) follows the highest column density material detected in previous {\it Herschel} (Fig.~\ref{fig:EMERGE_maps}a) and {\it WISE} (Fig.~\ref{fig:EMERGE_maps}b) observations. As expected, the low-resolution (30") IRAM-30m observations show prominent emission of N$_2$H$^+$ coinciding with gas at  $N\mathrm{(H_2)}\gtrsim 20\times10^{21}$~cm$^{-2}$ in this cloud ($A_V\gtrsim20$~mag; Fig.~\ref{fig:EMERGE_maps}d) while the HNC emission extends further towards columns as low as $N\mathrm{(H_2)}\lesssim 3\times10^{21}$~cm$^{-2}$ (or few $A_V$; Fig.~\ref{fig:EMERGE_maps}e). The detection of these tracers corresponds to different temperature regimes (Fig.~\ref{fig:EMERGE_maps}c) where N$_2$H$^+$ is usually found at low temperatures ($T_K\lesssim$~15~K) while HNC samples lukewarm material ($T_K\sim$~20-30~K).
These two tracers are also sensitive to distinct turbulent regimes. In particular, while N$_2$H$^+$ in found under mostly sub-sonic gas conditions ($\sigma_{NT}/c_s\lesssim 1$), HNC usually traces supersonic material ($\sigma_{NT}/c_s\sim 3-4$) in our targets \citep{BonanomiPaperVI}\footnote{$\sigma_{NT}$ and $c_s$ refer to the non-thermal velocity dispersion and sound speed, respectively \citep{Hacar2022}.}.
Unlike the largely filtered out ALMA-alone continuum (Fig.~\ref{fig:EMERGE_maps}f),
the true nature of the above N$_2$H$^+$ (Fig.~\ref{fig:EMERGE_maps}g) and HNC (Fig.~\ref{fig:EMERGE_maps}h) emission becomes apparent when observed at the enhanced resolution (4.5") and sensitivity of their ALMA+IRAM-30m maps revealing the complex fibrous sub-structure of these star-forming region \citep[see also][]{SocciPaperIII} extending over both dense and diffuse components.
Similar maps for all other regions of interest as well as a full discussion of their emission properties can be found in Paper I.

\section{From molecular emission to gas column density}\label{sec:mol2H2}

Beyond the description of their emission properties (see above), it is fundamental to quantify the conversion between their observed line integrated intensities and the actual $N\mathrm{(H_2)}$ column densities in our maps in order to calculate key physical parameters such as gas column densities, radial profiles, and masses.
Compared to previous intensive methods including line emission modeling, in this section we aim to derive empirical conversion factors for the HNC (1-0) and N$_2$H$^+$ (1-0) lines by comparing the emission properties of these transitions with the corresponding total column density detected by {\it Herschel}.

Our choice of molecular tracers is motivated by the results of previous molecular surveys. The N$_2$H$^+$ (1-0) line is a density selective tracer of the dense gas above $n(\mathrm{H_2)>10^4~cm}^{-2}$ found at gas column densities above $N(\mathrm{H_2})>10^{22}~\mathrm{cm}^{-2}$ in molecular clouds with observational properties are well documented in the literature \citep{Pety2017,Kauffmann2017,Tafalla2021}.
On the other hand, the HNC (1-0) line intensity exhibits a close-to linear dependence with the total gas column density \citep[Pearson coefficient $p=0.90$;][]{Tafalla2021,Tafalla2023}\footnote{We remark that the integrated HNC (1-0) emission shows the strongest correlation with column density (highest p-value) of all traditional 3mm line tracers, namely, CO and its isotopologues, HCO$^+$, HCN, HNC, CS, and N$_2$H$^+$ (1-0) lines according to \citet{Tafalla2021,Tafalla2023}.} becoming a reliable proxy of gas distribution of the gas at intermediate column densities $N(\mathrm{H_2})\sim (3-100)\times10^{21}~\mathrm{cm}^{-2}$ \citep[see also][ for a full discussion]{BonanomiPaperVI}.
In fact, HNC appears as major contributor after C$^{18}$O in the retrieval of the total gas column density in multi-line studies at intermediate column densities \citep{Gratier2021}.
Compared to the latter, however, the comparatively weaker HNC emission requires observations of higher sensitivity.
These choices are also tailored to the column density regimes covered by our ALMA+IRAM-30m maps \citep[i.e. $N(\mathrm{H_2)>5\times10^{21}~\mathrm{cm^{-2}}})$; see][]{Hacar2024}. Extension of this analysis into lower column density regimes might require additional low column density tracers (e.g. $^{13}$CO).

In our analysis we adopt the simplest form for the RTE 
for a single, homogeneous and plane-parallel gas slab
in which the observed line intensity $T_{\mathrm{mb,\nu}}$ follows
\begin{equation}\label{eq:RTE}
    T_{\mathrm{mb,\nu}}=\bigl[J_\nu(T_{\mathrm{ex}})-J_\nu(T_{\mathrm{bg}})\bigr]\,(1-e^{-\tau_\nu}),
\end{equation}
where $J_\nu(T)=\frac{h\nu/k}{\exp(h\nu/kT)-1}$ is the Planck function. 
Our approach thus assumes homogeneous density and temperature gas conditions as well as constant molecular abundance along the line-of-sight. Likely deviating from the true gas conditions in our clouds, this simplified treatment permits us to derive robust conversion factors from observations to be applied at different resolutions (Sects.~\ref{sec:Xfactors}-\ref{sec:Xfactors_highres}). 
The caveats and limitations of such approximation are discussed in Sect.~\ref{sec:caveats}.

\subsection{Temperature-corrected, conversion factors at low resolution}\label{sec:Xfactors}


We start our analysis at the single-dish resolution. 
Our empirical derivations should consider two global properties systematically observed in our large-scale maps shown in Figs.\ref{fig:EMERGE_maps} a, c, d, and e. First and foremost, the integrated intensity of both N$_2$H$^+$ (1-0) and HNC (1-0) lines show a direct correlation with the gas column density $N\mathrm{(H_2)}$ detected in our {\it Herschel} maps. Second, and particularly prominent in regions such as OMC-1, the emission of both tracers becomes significantly brighter in regions with higher $T_K$ temperatures. The integrated line intensity $I(X)$ (HNC or N$_2$H$^+$) is expected to vary with the gas density $n(\mathrm{H}_2)$, the gas kinetic temperature $T_K$, the molecular abundance $X_i$, and therefore indirectly with $N\mathrm{(H_2)}$, according to the RTE (see Eq.~\ref{eq:RTE}).
In particular, an enhancement of the integrated line intensity $I(X)$ is predicted for increasing $T_K$ and $N\mathrm{(H_2)}$ values as they boost $T_{ex}$ and $\tau_{\nu}$, respectively. These dependence have been reported in previous observations in Orion \citep[e.g.,][]{Ungerechts1997,Hacar2020,Tafalla2023}, among others.

\begin{figure*}[ht!]
\centering
\includegraphics[width=0.9\textwidth]{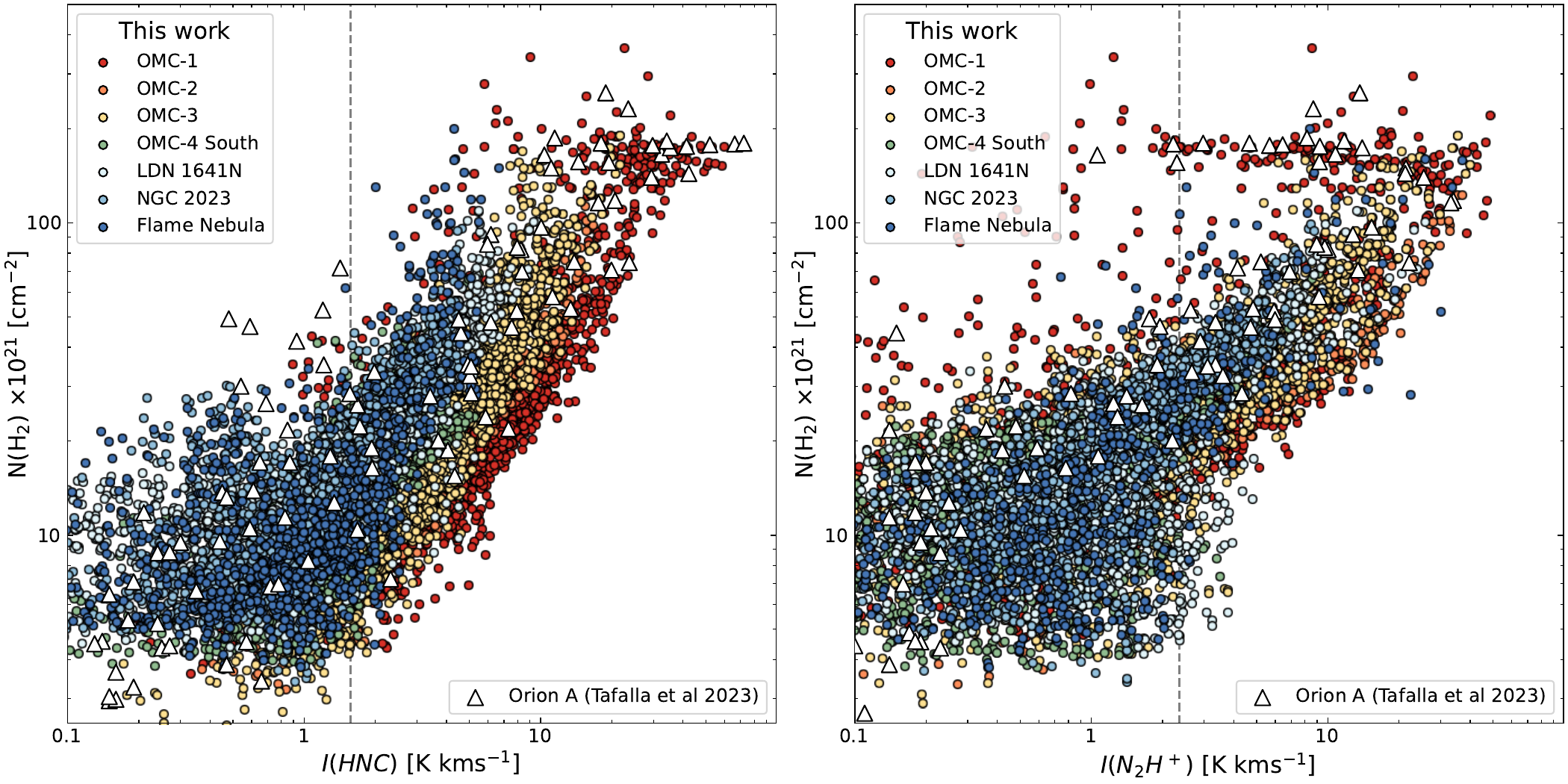}\\
\centering
\includegraphics[width=0.9\textwidth]{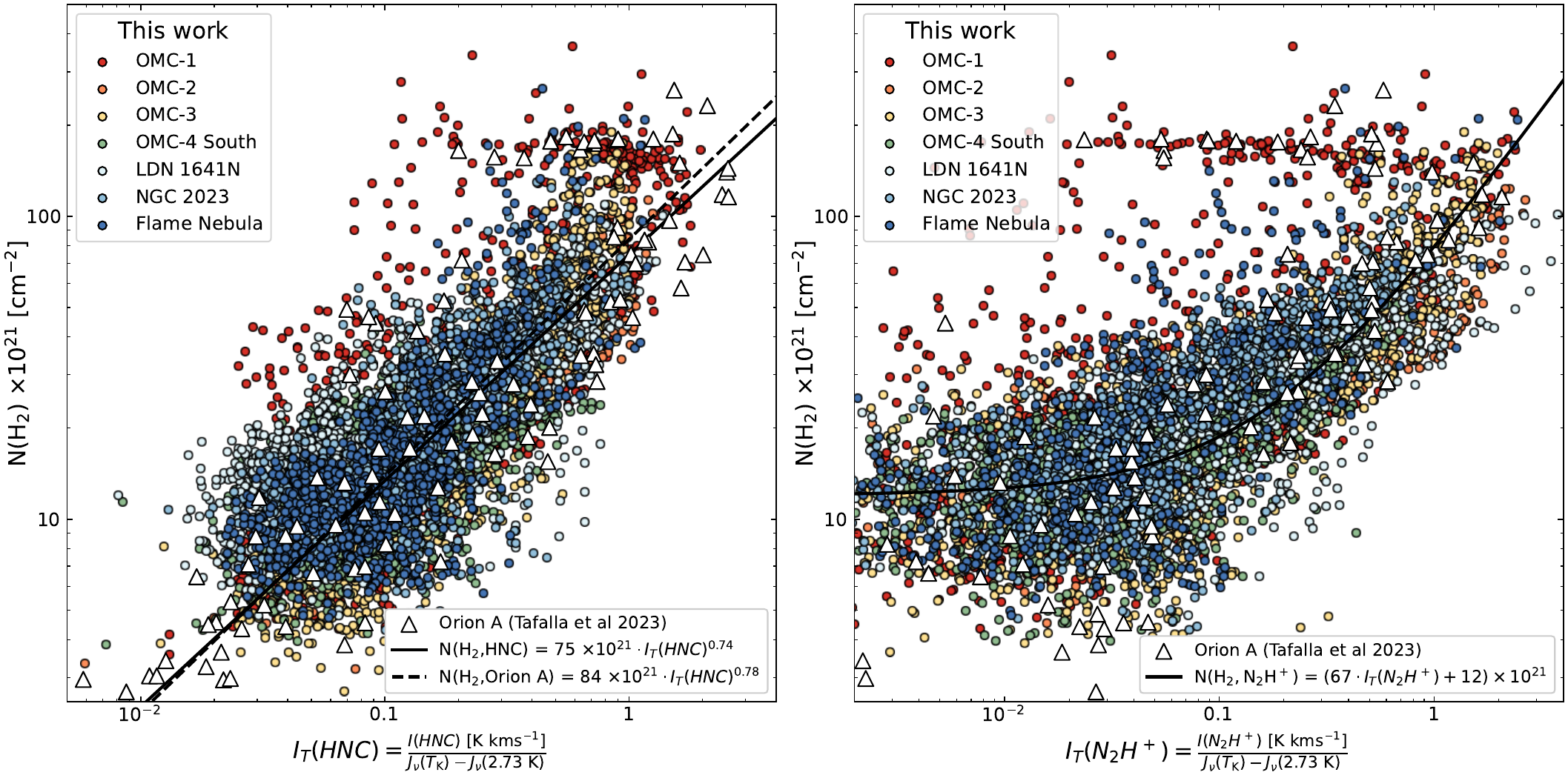}
      \caption{Pixel-by-pixel comparison of the {\bf (Left panels)} HNC (1-0) and {\bf (Right panels)} N$_2$H$^+$ (1-0) emission with the total $N\mathrm{(H_2)}$ column density detected at low-resolution in all regions in our sample (see color code in legend):
      {\bf (Top panels)} Comparisons between integrated intensity $I(X)$ and the {\it Herschel} gas column densities $N\mathrm{(H_2)}$;
      {\bf (Bottom panels)} Correlation between their integrated intensity normalized by temperature (i.e. temperature-corrected) $I_T(X)$ and the {\it Herschel} gas column densities $N(\mathrm{H_2})$.
      For comparison, we include the observed values reported by \citet{Tafalla2023} in Orion A in all our panels (white triangles).
      The value of the first intensity contours shown in Fig.\ref{fig:EMERGE_maps} are indicated in the top panels (vertical, grey dashed lines).
      The correlations described in Eqs. \ref{eq:X_HNC} and \ref{eq:X_N2H+} are displayed in the bottom panels (black lines).
      }
\label{fig:Xfactors}
\end{figure*}

To quantify these radiative transfer effects, we first display a pixel-by-pixel comparison between the I(HNC) and I(N$_2$H$^+$) integrated intensities and the {\it Herschel} gas column densities $N\mathrm{(H_2)}$ in Fig.~\ref{fig:Xfactors} (top panels). We observe a direct and strong correlation between the integrated emission of both HNC (top left panel) and N$_2$H$^+$ (top right panel) and the total gas column density in the seven regions included in the EMERGE Early ALMA Survey (color coded; see legend). This correlation holds between $N\mathrm{(H_2)}\sim 7\times10^{21}$~cm$^{-2}$ and $N\mathrm{(H_2)}\sim 150\times10^{21}$~cm$^{-2}$, and describe most of the positions in our maps. The only exception to this correlation are those OMC-1 positions in the vicinity of the Ori BN/KL region with $N\mathrm{(H_2)}> 150\times10^{21}$~cm$^{-2}$ (red points) known to present a particular chemistry \citep[see a detailed discussion in][]{Tafalla2023}. 

Yet, systematic variations also become apparent in this Fig.~\ref{fig:Xfactors} (top panels). While sharing a global trend, points from different regions appear to have slightly different slopes and offsets due to the effects of temperature. These temperature variations are particularly pronounced in the case of HNC which  traces more diffuse regions subject of larger temperature gradients.
N$_2$H$^+$, on the other hand, suffers less from these effects as it usually samples much colder regions with less temperature variations (see Fig.~\ref{fig:EMERGE_maps}c). Two cases illustrate these changes. The high temperatures seen across the entire OMC-1 region (red points) shift its points to the rightmost part of these diagrams. The case of the Flame Nebula (dark blue points) combines both diffuse and warm gas with dense and cold regions. As a result, the temperature effects at low $N\mathrm{(H_2)}$ produce a steeper dependence of the line intensities compared to other regions. Considering these temperature changes is therefore crucial to characterize the local line emission in our targets.

Using Fig.~\ref{fig:Xfactors} (top panels) we could in principle derive specific correlations (X-factors) between $I$(X) and $N\mathrm{(H_2)}$ in each individual region, although this approach is far from optimal. 
Instead, we correct the above temperature dependence using the properties of the RTE in Eq.~\ref{eq:RTE}.
Following an analogous approach to \citet{Tafalla2023}, we divide the observed line intensity by its corresponding excitation term in Local Thermodynamical Equilibrium (LTE; with $T_{ex}=T_{K}$):
\begin{equation}\label{eq:I_norm}
    I_T(X)=\frac{I(X)}{\left(J_{\nu}(\mathrm{T}_{K})-J_{\nu}(\mathrm{T}_{bg})\right)}= \int \left(1-e^{-\tau_{\nu}}\right)\cdot d\nu,
\end{equation}
where $T_{K}$ is the gas kinetic temperature. 
With this normalization, we aim to equalize the observe temperature variations and characterize those variations that solely depend on $N\mathrm{(H_2)}$.

We display the comparison between the normalize $I_T(X)$ intensities and $N\mathrm{(H_2)}$ in Figure~\ref{fig:Xfactors} (bottom panels) for all pixels within our IRAM-30m maps (e.g., see Fig.~\ref{fig:EMERGE_maps}). The temperature normalization in Eq.~\ref{eq:I_norm} produces significant improvements on the correlations of both HNC (bottom left panel) and N$_2$H$^+$ (bottom right panel). After the removal of the temperature variations $I_T$(HNC) and $I_T$(N$_2$H$^+$) show uniform distributions with $N\mathrm{(H_2)}$ with no systematic difference between clouds. The only obvious exceptions are again Ori BN/KL 
 and (part of) Flame Nebula
which clearly deviate from these trends\footnote{We notice that the temperature normalization in Ori BN/KL enhances the deviations from the average $I_T(\mathrm{X})$ correlation due to the high temperatures found in this region \citep[$T_K>$~50~K; e.g.,][]{Hacar2020}.}.
Outside these peculiar regions all clouds in our sample show a tight and homogeneous dependence of $I_T$(HNC) and $I_T$(N$_2$H$^+$) with $N\mathrm{(H_2)}$ for column densities up to $N\mathrm{(H_2)}\sim 100\times10^{21}$~cm$^{-2}$. Still, each molecule shows a distinct dependence with $N\mathrm{(H_2)}$.
The variations of $I_T$(HNC) seems to follow a quasi-linear relation with $N\mathrm{(H_2)}$ for the entire range of column densities sampled by this molecule. On the other hand, the dependence of $I_T$(N$_2$H$^+$) is much shallower and shows a cut-off at around $N\mathrm{(H_2)}\sim 10\times10^{21}$~cm$^{-2}$ (or $\sim$10~A$_V$) set by the effective threshold for CO depletion \citep{Tafalla2021}.

Similar to the estimators used in machine-learning studies \citep[e.g.,][]{Gratier2021,Shimajiri2023_H2}, we have quantified the improvement of our temperature corrections by calculating the Root Mean Square Logarithmic Error (RMSLE) in our HNC (all) combined data with respect to a linear fit (in log-space) as:
\begin{equation}\label{eq:RMSLE}
    RMSLE =\sqrt{\frac{1}{n}\sum_{i=1}^n\left(log\ \hat{y_i}-log\ y_i \right)^2},
\end{equation}
where $y_i$ is the observed value of $\mathrm{N(H_2)}$ for a given pixel $i$ and $\hat{y}_i$ to the expected value given the linear fit of the corresponding $I(\mathrm{HNC})$ or $I_T(\mathrm{HNC})$ emission in Figs.~\ref{fig:Xfactors} (left panels), respectively. 
The temperature normalization reduces the RMSLE from 0.96 to 0.18~dex, that is a factor of $\sim$~5 reduction in linear space.
On the other hand, the RMSLE compared to a linear fit of our data is only marginally reduced from 0.22 to 0.16~dex (factor $\sim 1.3$) in the case of $I(\mathrm{N_2H^+})$ and $I_T(\mathrm{N_2H^+})$ (right panels), respectively.

In the light of the results in Fig.~\ref{fig:Xfactors} (lower panels) we derived unique temperature-corrected,  conversion factors (i.e. $I_T(\mathrm{X})$ vs $N\mathrm{(H_2)}$ for HNC and N$_2$H$^+$. 
Given the linear dependence seen for HNC (lower left panel), we obtain a linear fit (in log-log space; black solid line) combining all the points in Fig.~\ref{fig:Xfactors} (lower left panel) with $N\mathrm{(H_2)}\leq 100\times10^{21}$~cm$^{-2}$ leading to a $I_T(\mathrm{HNC})$ to $N\mathrm{(H_2)}$ conversion factor for the diffuse gas such as:
\begin{equation}\label{eq:X_HNC}
    N_\mathrm{{HNC}}(\mathrm{H_2})~[\mathrm{cm}^{-2}] = 75\times 10^{21} \cdot \left(\frac{I_T(\mathrm{HNC})}{[\mathrm{K~km~s}^{-1}]}\right)^{0.74}~.
\end{equation}
Similar to \citet{Tafalla2023}, this quasi-linear dependence of $I_T(\mathrm{HNC})$ highlights the use of the HNC(1-0) line as robust tracer of the cloud material down to $N_\mathrm{{HNC}}\mathrm{(H_2)}\lesssim (5-8)\times10^{21}$~cm$^{-2}$ \citep[or 5-8 A$_V$; see also][]{Gratier2021}. Despite its good behavior across a large range of column densities, we note a deviation from this relation at $N_\mathrm{{HNC}}\mathrm{(H_2)}\gtrsim 50-60\times10^{21}$~cm$^{-2}$ where Eq.\ref{eq:X_HNC} seems to underestimate the observed $N\mathrm{(H_2)}$ values.

The analysis of $I_T(\mathrm{N}_2\mathrm{H}^+)$ is complicated by depletion. As shown in Fig.~\ref{fig:Xfactors} (lower right panel), $I_T(\mathrm{N}_2\mathrm{H}^+)$ shows a linear variation at high column densities above a certain column density threshold at few 10$^{22}$~cm$^{-2}$. The dependence of $I_T(\mathrm{N}_2\mathrm{H}^+)$ however flattens at lower column densities. \citet{Hacar2018} provided an empirical fit to the $I_T(\mathrm{N}_2\mathrm{H}^+)$ variations in OMC-1 and OMC-2 following\footnote{Strictly speaking \citet{Hacar2018} normalized their values {\bf with} respect to $T_K$ only instead of the Planck function used in Eq.~\ref{eq:X_N2H+}. We notice that for the frequency and temperature regimes explored in this work $\left(J_{\nu}(\mathrm{T}_{K})-J_{\nu}(\mathrm{T}_{bg})\right) \sim \mathrm{T}_K$, the changes within these two normalizations are within the noise of our data.}:
\begin{equation}\label{eq:X_N2H+}
    N_\mathrm{{N_2H^+}}(\mathrm{H_2})~[\mathrm{cm}^{-2}] = \left( 67.4  \times \left[\frac{I_T(\mathrm{N}_2\mathrm{H}^+)}{[\mathrm{K~km~s}^{-1}]}\right] + 12\right) \times 10^{21},
\end{equation}
obtained form the combination of a linear variation proportional to $67.4\cdot 10^{21}$~cm$^{-2}$ and a column density threshold of $12.0\cdot 10^{21}$~cm$^{-2}$. 
While approximated, this result satisfactorily reproduces the behavior observed in other regions for column densities up to $N\mathrm{(H_2)}\sim 100-200\times10^{21}$~cm$^{-2}$. The conversion described by Eq.~\ref{eq:X_N2H+} has been successfully implemented in previous papers of this series for the study of $\mathrm{N_2H^+}$ \citep{SocciPaperIII,SocciPaperIV}.
Similar conversion factors for $\mathrm{N_2H^+}$ are also found in young proto-clusters \citep{Peretto2023,Anderson2026}.
We thus have adopted this last relationship as description of our $I_T(\mathrm{N}_2\mathrm{H}^+)$ data points. Compared to $I_T(\mathrm{HNC})$, we observe a tighter correlation between $I_T(\mathrm{N}_2\mathrm{H}^+)$ and $N\mathrm{(H_2,N_2H^+)}$ at high column density values becoming a better descriptor of the cloud material at  $N(\mathrm{H_2})\gtrsim 60\times10^{21}$~cm$^{-2}$ (or $>60$ A$_V$) where HNC is limited by depletion.

Equations \ref{eq:X_HNC} and \ref{eq:X_N2H+} provide empirical but direct conversion factors between the observed line intensities plus gas temperatures and the total gas column densities in our maps.
Among the most surprising things is the robustness of these correlations independently of the physical conditions of our targets (see Paper I). Outside the Ori BN/KL region, these conversion factors work out across more than one order of magnitude range in $N(\mathrm{H_2})$ with only minor deviations of $\sim$~0.2-0.3~dex.
Remarkably, the observed correlations reproduce not only our data (colored points) but also the average properties derived for the entire Orion A and appear to continue towards lower $N\mathrm{(H_2)}$ values below our detection threshold \citep[white triangles; see][]{Tafalla2023}. A linear fit of Tafalla et al data (black dashed line) within the overlapping range of column densities, i.e. $[5,100]\times10^{21}$~cm$^{-2}$, show the similarity between these results.

\subsection{Molecular conversion factors at high-resolution}\label{sec:Xfactors_highres}

\begin{figure*}[t]
\centering
\includegraphics[width=1.\textwidth]{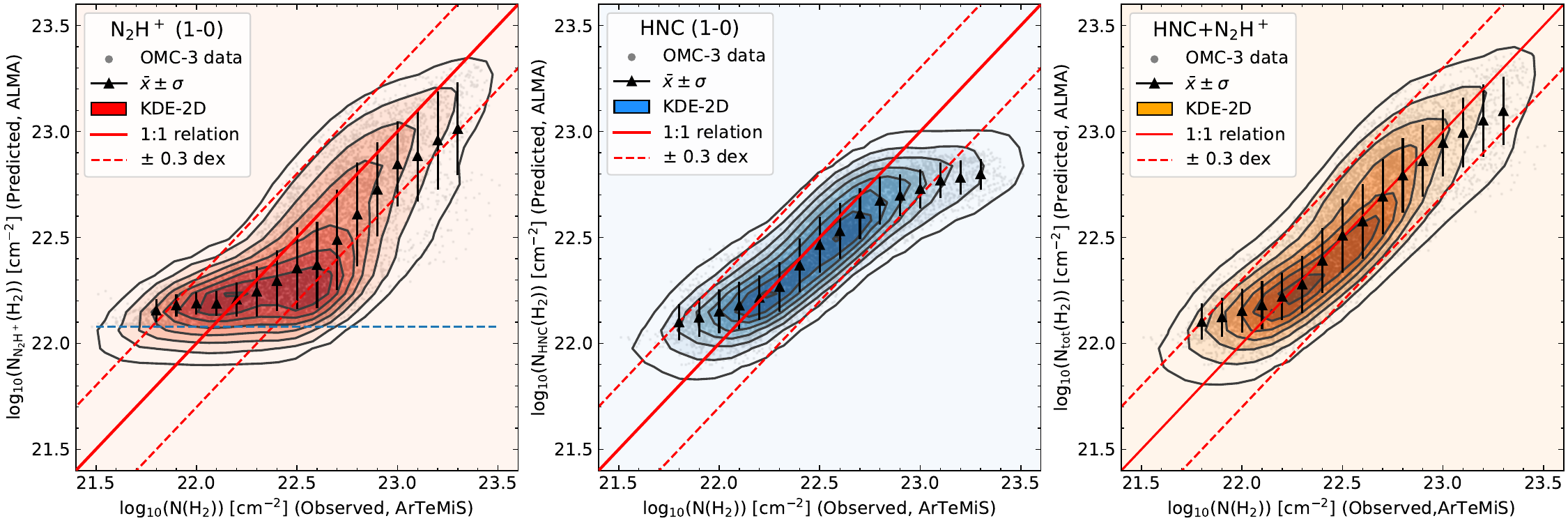}
      \caption{Pixel-by-pixel (grey dots), and their corresponding 2D kernel density estimate (KDE-2D; coloured areas), comparison between the total gas column density $N\mathrm{(H_2)}$ observed with ArTeMiS+Herschel \citep{Schuller2021} and our predicted values obtained using Eqs.~\ref{eq:X_HNC} and \ref{eq:X_N2H+} at 8" resolution.
      From left to right: 
      {\bf (Left panel)} Predicted $N_\mathrm{{N_2H^+}}(\mathrm{H_2)}$ values for the dense gas using N$_2$H$^+$ and Eq.~\ref{eq:X_N2H+};
      {\bf (Central panel)} Predicted $N_\mathrm{{HNC}}(\mathrm{H_2)}$ values for the diffuse gas using HNC and Eqs.~\ref{eq:X_HNC};
      {\bf (Right panel)} Predicted $N_\mathrm{tot}(\mathrm{H_2)}$ obtained from the combination of both diffuse and dense components using Eq.\ref{eq:X_total}.
      We indicate the 1:1 correlation line (red solid line) as well as deviations of $\pm 0.3$~dex from it (aka factor 2; red dashed lines) in all panels. We mark the column density floor for $N_\mathrm{{N_2H^+}}(\mathrm{H_2)}$ (=$12\times 10^{21}$~cm$^{-2}$, see Eq.~\ref{eq:X_N2H+}) in the central panel (horizontal dashed line). 
      In all panels, we display the mean and dispersion values ($\hat{x}\pm \sigma$; black triangle plus segment) of the predicted $N_i$ column density binned every 0.1~dex in column density as observed by ArTeMiS.
      }
\label{fig:Xfactors_highres}
\end{figure*}

Equations \ref{eq:X_HNC} and \ref{eq:X_N2H+} could become a powerful tool to investigate the mass distribution of both diffuse and dense components in our clouds, particularly if these conversion factors could be applied at different resolutions. 
It is therefore essential to explore whether our molecular calibrations obtained at the IRAM-30m resolution (30") can be exported into our high-resolution ALMA data (4.5").  The lack of short-spacing information in our 3mm-continuum data prevents a direct comparison at the native ALMA resolution (Fig.~\ref{fig:EMERGE_maps}f). Interestingly, the OMC-3 region has been surveyed by \citet{Schuller2021} combining continuum APEX-ArTeMiS 350+450$\mu$m with {\it Herschel}-SPIRE (hereafter ArTeMiS for simplicity) observations. Using these maps, Schuller et al obtained independent measurements of the total $N\mathrm{(H_2)}$ column density across this cloud at 8" resolution. With a four times smaller beam than {\it Herschel} (see Fig.\ref{fig:EMERGE_maps}a), these observations become an excellent benchmark to test the validity of Eqs.~\ref{eq:X_HNC} and \ref{eq:X_N2H+} to recover the H$_2$ gas column density at resolutions comparable to ALMA.

To carry out these comparisons we first convolved our ALMA+IRAM-30m (MACF) HNC and N$_2$H$^+$ maps into a final resolution of 8", similar to the ArTeMiS maps. We then converted these molecular maps into their corresponding column densities using Eq.\ref{eq:X_HNC} and Eq.\ref{eq:X_N2H+}, respectively, using the nearest $T_K$ value obtained in our IRAM-30m observations (at 30") for normalization. After regridding them into a common frame, in Fig.\ref{fig:Xfactors_highres} we display a pixel-by-pixel comparison between the predicted $N_\mathrm{{N_2H^+}}\mathrm{(H_2)}$ (left panel) and  $N_\mathrm{{HNC}}\mathrm{(H_2)}$ (central panel) values and the observed  $N\mathrm{(H_2)}$ column densities observed in continuum by \citet{Schuller2021}. Given the large number of pixels in these images, we also added the corresponding 2D kernel density estimates (KDE-2D; colored areas). 

The results of these comparisons are consistent with those obtained at the single-dish resolution (Sect.\ref{sec:Xfactors}).
We overplot the mean value and dispersion ($\hat{x}\pm\sigma$) of our $N_\mathrm{{N_2H^+}}\mathrm{(H_2)}$ and $N_\mathrm{{HNC}}\mathrm{(H_2)}$ estimates binned every 0.1~dex in Fig.\ref{fig:Xfactors_highres} (black triangles and error bars).
Our $N_\mathrm{{HNC}}\mathrm{(H_2)}$ measurements (central panel)  reproduce the column densities obtained by these ArTeMiS observations within (on average) less than a factor of 2 difference ($\pm$~0.3~dex) up to column densities of $\sim 60\times 10^{21}$~cm$^{-2}$ where the effects of HNC depletion limits the use of this tracer. On the other hand, our $N_\mathrm{{N_2H^+}}\mathrm{(H_2)}$ estimates  (left panel) show better agreement when describing higher column density values up to few $10^{23}$~cm$^{-2}$.

Given these complementary results, a combination of N$_2$H$^+$ and HNC measurements is thus needed to trace the total gas column density $N_\mathrm{tot}(H_2)$ in our ALMA maps. Ideally, this combination should take into account the increasing contribution of the dense gas traced by N$_2$H$^+$ at high column densities while HNC is progressively depleted. This combination should also take into account the superposition of both low and high column density components along the line of sight in regions where both molecules were detected. To model these effects, we add the contributions of both HNC and $\mathrm{N_2H^+}$ gas components adopting a simple weighting scheme such as:
\begin{equation}\label{eq:X_total}
    N_\mathrm{{tot}}(\mathrm{H}_2)= N_\mathrm{{HNC}}(\mathrm{H_2})+ N_\mathrm{{N_2H^+}}(\mathrm{H_2})\times \mathcal{W}(\mathrm{N_2H^+}),
\end{equation}
where $N_\mathrm{{HNC}}(H_2)$ and $N_\mathrm{{N_2H^+}}(H_2)$ are again defined by Eq.\ref{eq:X_HNC} and Eq.\ref{eq:X_N2H+}, respectively, and the transfer function $\mathcal{W}\mathrm{(N_2H^+)}$ for $\mathrm{N_2H^+}$. 
In this context, we aim for a $\mathcal{W}\mathrm{(N_2H^+)}$ function that describes a smooth transition between the low (HNC) and high column density ($\mathrm{N_2H^+}$) regimes in our clouds and that accounts for their combined contribution when superposed along the line-of-sight. We thus adopt $\mathcal{W}\mathrm{(N_2H^+)}$ to take the shape of a logistic function (i.e. sigmoid transfer function; $\mathcal{W}(x)=\frac{1}{1+exp(-k(x-x_0))}$ with $\mathcal{W}\in[0,1]$) regularly employed as activation function in other astronomical and statistical problems.
Assuming $x-x_0=log_{10}\left(I_T(\mathrm{N}_2\mathrm{H}^+)\right)$ (i.e. $\propto log_{10}(N_\mathrm{N_2H^+}\mathrm{(H_2)})$ following Eq.\ref{eq:X_N2H+}), a series of empirical tests suggest the use of a logistic growth $k=2$, that is
\begin{equation}\label{eq:X_totalW}  
    \mathcal{W}(\mathrm{N_2H^+}) = \frac{1}{1+e^{\left( -2 \cdot log_{10}I_T(\mathrm{N}_2\mathrm{H}^+)\right)}}=\frac{1}{1+I_T(\mathrm{N}_2\mathrm{H}^+)^{\left(\frac{-2}{ln(10)}\right)}}.
\end{equation}
Our choice of parameters for the logistic function $\mathcal{W}\mathrm{(N_2H^+)}$ makes $N_\mathrm{tot}\mathrm{(H_2)}$ to be primarily determined by the diffuse component $N_\mathrm{HNC}\mathrm{(H_2)}$ with only increasing contributions of the dense gas $N_\mathrm{N_2H^+}\mathrm{(H_2)}$ ($>$50\%) at column densities of $\gtrsim 70\times 10^{21}$~cm$^{-2}$, in agreement to properties reported in Sect.\ref{sec:Xfactors}.

We show the predicted values for $N_\mathrm{tot}(H_2)$ using Eq.\ref{eq:X_total} in Fig.\ref{fig:Xfactors_highres} (right panel). The combined contributions of the diffuse and dense components extends the dynamic range of column densities recovered in our data showing a direct and consistent correspondence with the previous ArTeMiS results across almost two orders of magnitude between $N_\mathrm{tot}\mathrm{(H_2)}\simeq 10^{21.8}-10^{23.5}$ cm$^{-2}$. Remarkably, our $N_\mathrm{tot}\mathrm{(H_2)}$ estimates reproduce the ArTeMiS results typically within $\lesssim$~0.3~dex difference (red dashed lines; see also Sect.~\ref{sec:accuracy}, where these discrepancies are consistent with the intrinsic dispersion of our fits in Fig.\ref{fig:Xfactors} (lower panels).

\subsection{Conversion accuracy and comparison with previous results}\label{sec:accuracy}

\begin{figure*}[h]
\centering
\includegraphics[width=1.\textwidth]{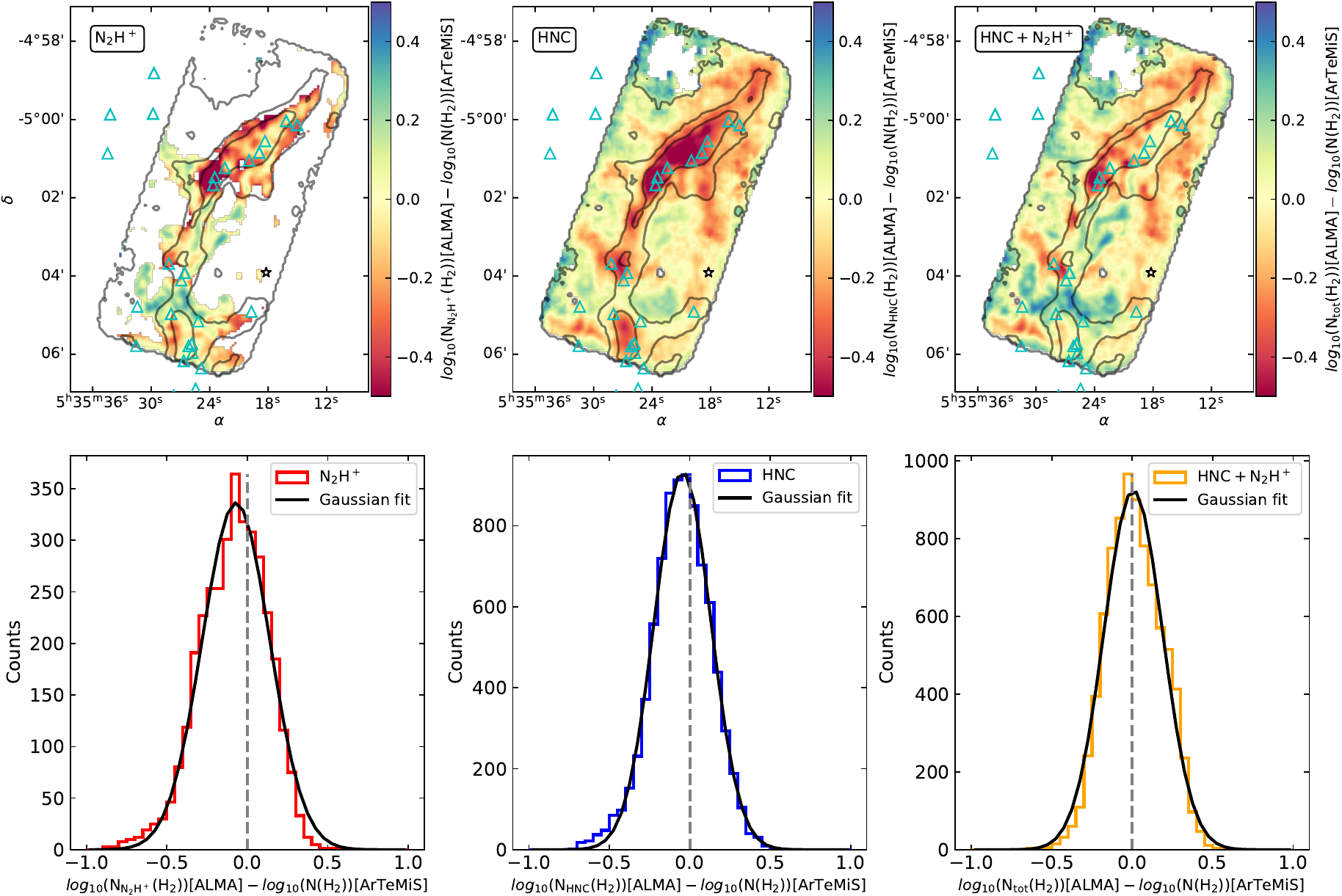}
      \caption{Logarithmic deviation (i.e. $log_{10}(N_i\mathrm{(H_2)[Pred.]})-log_{10}(N\mathrm{(H_2)[Observed]})$)
      between the predicted
      {\bf(left panels)} $N_\mathrm{N_2H^+}\mathrm{(H_2)}$, {\bf(central panels)} $N_\mathrm{HNC}\mathrm{(H_2)}$, and
      {\bf(right panel)} $N_\mathrm{tot}\mathrm{(H_2)}$ ALMA column density estimates and the observed $N\mathrm{(H_2)}$ reported by ArTeMiS at 8'' resolution in OMC-3.
      {\bf (Top panels)} Pixel-by-pixel distribution, i.e. logarithmic error maps. For comparison all panels include the ArTeMiS contours at $N\mathrm{(H_2)}=[10,50,100]\times10^{21}~\mathrm{cm^{-1}}$ (gray contours). Symbols are similar to those in Fig.~\ref{fig:EMERGE_maps}.
      {\bf (Bottom panels)} Histograms.
      Only pixels with $I\mathrm{(\mathrm{N_2H^+})\ge 2.5~K~km~s^{-1}}$ and $I\mathrm{(\mathrm{HNC})\ge 1.5~K~km~s^{-1}}$ (see Sect.~\ref{sec:EMERGE}) are included in this plots. The histograms in the lower panel include the best Gaussian fit to the data (black solid line).
      }
\label{fig:differences}
\end{figure*}

The above combination of low, $N_\mathrm{HNC}\mathrm{(H_2)}$, and high, $N_\mathrm{N_2H^+}\mathrm{(H_2)}$, column density components into a total $N_\mathrm{tot}\mathrm{(H_2)}$ estimate using Eq.~\ref{eq:X_totalW} also improves the accuracy of our total column density measurements in OMC-3.
A pixel-by-pixel comparison with the ArTeMiS column density values of these three estimates indicate that the RMSLE (Eq.~\ref{eq:RMSLE}) is reduced from 0.25~dex and 0.19~dex in the case of $N_\mathrm{N_2H^+}\mathrm{(H_2)}$ and $N_\mathrm{HNC}\mathrm{(H_2)}$. respectively, down to 0.17~dex for $N_\mathrm{tot}\mathrm{(H_2)}$.

Although significant over a wide range of column densities this improvement on the column density estimate is not uniform. As seen from the mean and dispersion values per column density bin (black triangles plus error bars in Fig.\ref{fig:Xfactors_highres}, right panel), the differences between the observed (ArTeMiS) and predicted (ALMA) column density values increase outside the optimal sensitivity range of our selected tracers. In particular, we observed departures from the optimal correspondence below $\mathrm{N(H_2)\lesssim 10^{22}~cm^{-2}}$ (overestimated) and above $\mathrm{N(H_2)\gtrsim 10^{23}~cm^{-2}}$ (underestimated) inherited from the similar variations found in the $\mathrm{HNC}$ (central panel) and $\mathrm{N_2H^+}$ (left panel) components, respectively.

Following \citet{Gratier2021}, we further illustrate these comparisons in Fig.~\ref{fig:differences} from the analysis of the spatial variations of the logarithmic deviation  between our predicted column density estimates in each case and the observed column densities reported by ArTeMiS \citep{Schuller2021}, that is $log_{10}(N_i\mathrm{(H_2)[Pred.]})-log_{10}(N\mathrm{(H_2)[Observed]})$. Overall, $N_\mathrm{{N_2H^+}(H_2)}$ (left panel) satisfactorily reproduces ($\pm0.2$~dex) the observed column density above $N\mathrm{(H_2)> 10^{22}~cm^{-2}}$ (left panels), while $N_{\mathrm{HNC}}(H_2)$ (central panels) does it better below such threshold (second gray contour in upper panels). The combination of the two into $N_{tot}\mathrm{(H_2)}$ (right panels) further improves these correspondence leading to a more uniform recovery of the observed total column density recovered by ArTeMiS (see the fit of the corresponding distributions in the lower panels).  
Despite these general good response of our molecular calibrations, localized regions with embedded protostars, such as SMM-7 \citep[reddest spot at the center of OMC-3;][]{Takahashi2013}, remain challenging likely due to the feedback effects of YSOs into their surrounding molecular material. 

Our molecular calibration complement the results obtained by previous multi-line studies using machine learning techniques.
\citet{Gratier2021} introduced the use of {\it random forest} in hyperspectral line analysis including CO, HCO$^+$, HCN, HNC, and $\mathrm{N_2H^+}$ (and other 3~mm lines) to infer the total gas column density in Orion B. These works demonstrate that the use of multiple line tracers ($\geq 4$) can reduce the accuracy of the resulting RMSLE=0.09~dex. Similarly, \citet{Shimajiri2023_H2} combined {\it Extra Trees Regressor} for the analysis of the three main CO isotopologues to reproduce the column density distribution of gas along the ISF reaching an accuracy of RMSLE=0.16~dex. Our molecular conversion factors achieve similar accuracy (RMSLE=0.17~dex) when calibrated against high sensitivity single-dish observations and applied to high dynamic range ALMA+IRAM-30m data.

Critically, the previous works show optimal results at low and intermediate column densities ($\mathrm{N(H_2)\lesssim 15\times10^{21}~cm}^{-2}$) while the performance of these methods decreases at higher columns unless high density selective tracers such as $\mathrm{N_2H^+}$ are included in the analysis \citep[see][for a discussion]{Gratier2017}. 
This has been demonstrated in the case of OMC-3 by \citet{Shimajiri2023_H2} where the 
combination of $^{12}$CO, $^{13}$CO, and C$^{18}$O systematically underestimates the total column densities above $\mathrm{N(H_2)\gtrsim 2\times10^{22}~cm}^{-2}$ due to the strong depletion of CO in this region \citep[see Fig.5 in][]{SocciPaperIII}. 
In contrast, the combination of $\mathrm{HNC}$ and $\mathrm{N_2H^+}$ extends the dynamic range of these column density retrievals deeper into the cloud covering more opaque regions up to $\mathrm{N(H_2)\sim10^{23}~cm}^{-2}$ (see Fig.~\ref{fig:Xfactors_highres}).

\section{Obtaining high-dynamic range, column density maps with ALMA}\label{sec:NH2colmaps}

Despite being calibrated at lower resolution (30"), the results seen in Fig.~\ref{fig:Xfactors_highres} illustrate the robustness and predictive power of our molecular conversion factors.
The excellent agreement between our convolved ALMA maps (this work) and the ArTeMiS observations \citep{Schuller2021}
also supports the use of Eqs.~\ref{eq:X_HNC}-\ref{eq:X_total} as proxy of the total H$_2$ column density at high spatial resolutions with similar (or at least comparable) accuracy ($\pm$~0.2~dex) to those obtained in continuum by ArTeMiS.
This section explores the use of these calibrations to derive high dynamic range column density maps using high resolution ALMA molecular observations.

\subsection{Column density maps at 2000~au resolution}\label{sec:NH2maps}

\begin{figure*}[h]
\centering
\includegraphics[width=1.\textwidth]{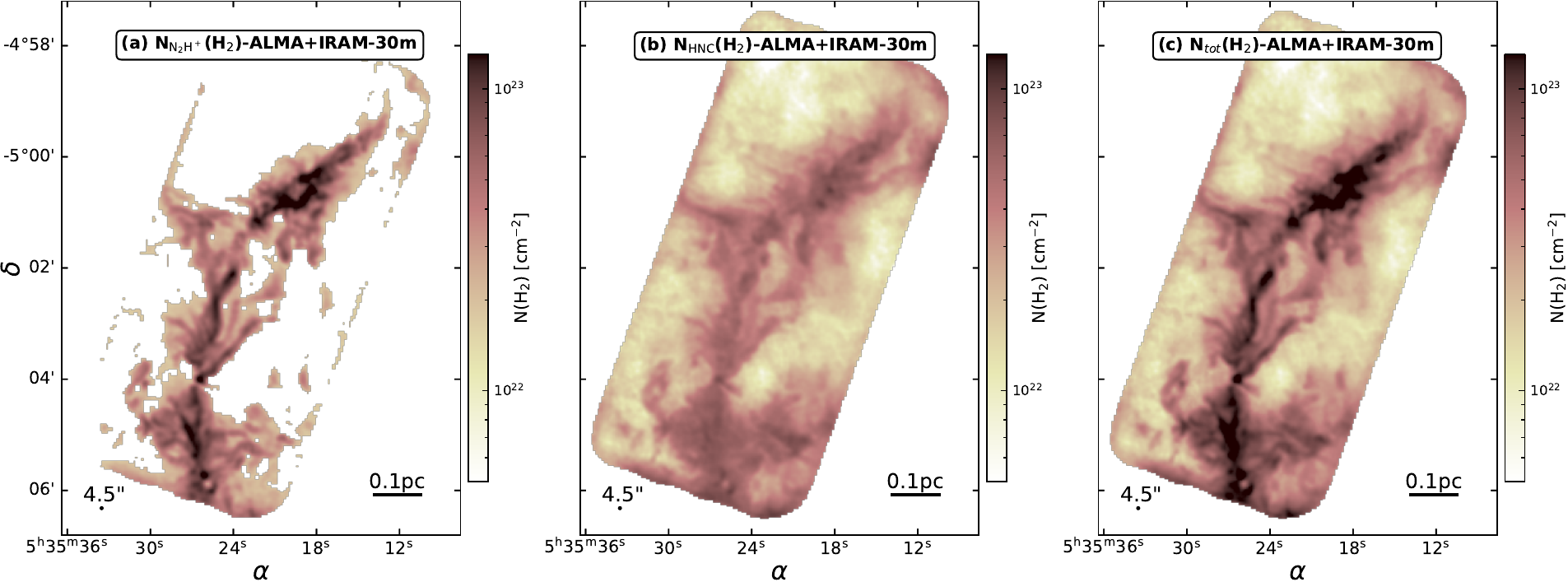}
      \caption{$N\mathrm{(H_2)}$ column density estimates in OMC-3. From left to right and from top to bottom: 
      {\bf (a)}$N_\mathrm{{N_2H^+}}(\mathrm{H_2)}$ (Eq.\ref{eq:X_N2H+}; only positions with $\ge$~1~K~km~s$^{-1}$ are shown), 
      {\bf (b)} $N_\mathrm{{HNC}}(\mathrm{H_2)}$ (Eq.\ref{eq:X_HNC}), and 
      {\bf (c)} $N_\mathrm{{tot}}(\mathrm{H_2)}$ (Eq.\ref{eq:X_total}) maps obtained from our ALMA+IRAM-30m MACF observations (4.5"). Symbols are similar to those in Fig.~\ref{fig:EMERGE_maps}. To facilitate their comparison, all maps are displayed within the same column density range. 
      Similar plots for all the other regions can be found in Appendix~\ref{appendix:DP}.
      }
\label{fig:OMC3_NH2}
\end{figure*}

We have created individual $N_\mathrm{{N_2H^+}}\mathrm{(H_2)}$, $N_\mathrm{{HNC}}\mathrm{(H_2)}$, and $N_\mathrm{{tot}}\mathrm{(H_2)}$ column density maps applying Eqs.~\ref{eq:X_HNC}-\ref{eq:X_total} to our ALMA molecular maps (4.5"). We compute $N_\mathrm{tot}(\mathrm{H_2)}$ for all targets including ALMA N$_2$H$^+$ and HNC data (OMC-3, OMC-4~South, LDN~1641N, NGC~2023, and Flame Nebula) and where $T_K$ information (at single-dish resolution) was available \footnote{We note that even when $T_K$ maps are available, individual positions with no reliable $T_K$ values are blanked (e.g., see Fig.\ref{fig:DP_OMC4_NH2}).}. In the case of $N_\mathrm{{N_2H^+}}\mathrm{(H_2)}$, we only consider positions with intensity $\ge$~1.5~K~km~s$^{-1}$ (i.e. $\gtrsim 20\times 10^{21}$~cm$^{-2}$ according to Eq.~\ref{eq:X_N2H+}). We display the results for OMC-3 in Fig.~\ref{fig:OMC3_NH2} while similar maps for OMC-4~South, LDN~1641N, NGC~2023, and Flame Nebula can be found in Figs.~\ref{fig:DP_OMC4_NH2}-\ref{fig:DP_FlameNeb_NH2} of Appendix~\ref{appendix:DP}. All column density maps shown in these figures are included in the EMERGE public DR4 accompanying this paper\footnote{See \url{https://emerge.univie.ac.at/results/data/}}.
Additional ALMA-12m observations of HNC (1-0) are needed to extend these results to the analysis of OMC-1 and OMC-2.

\begin{figure*}[t!]
\centering
\includegraphics[width=0.6\textwidth]{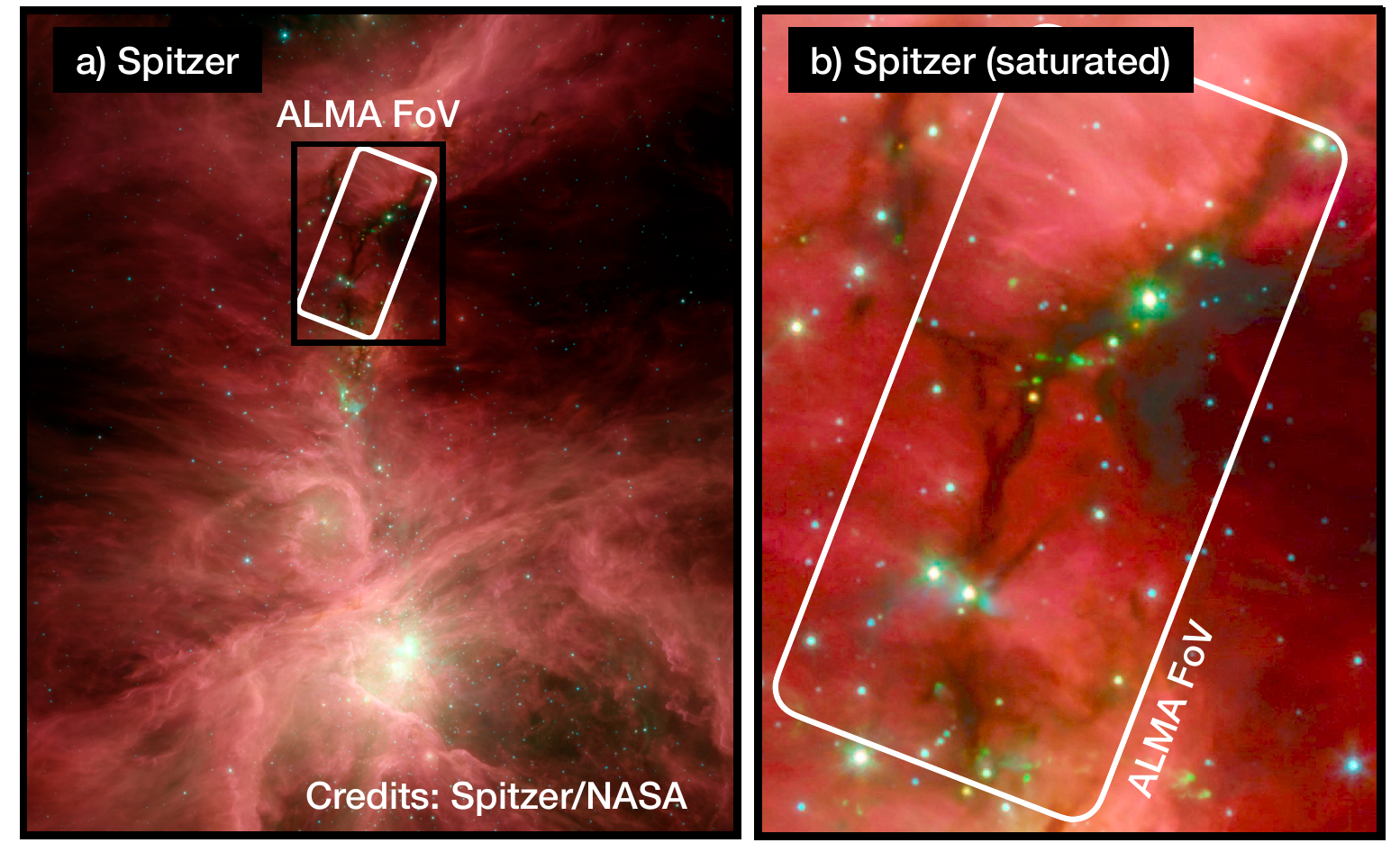}\\
\includegraphics[width=1.\textwidth]{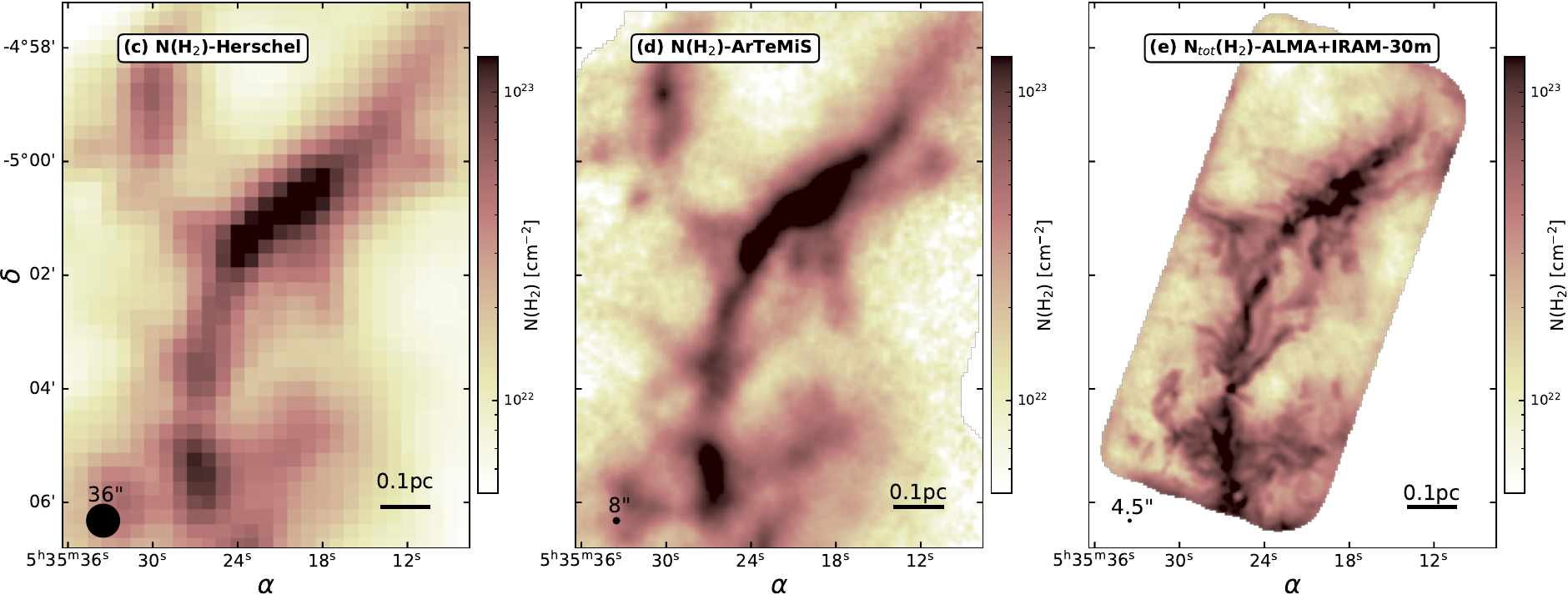}
      \caption{$N\mathrm{(H_2)}$ column density estimates in OMC-3 at different resolutions. From left to right and from top to bottom: 
      {\bf (a)} and {\bf (b)} {\it Spitzer} 3.6, 4.5, 5.8, and 8.0 $\mu m$ color-composed image \citep[Credits: {\it Spitzer}/NASA/JPL-Caltech; see also ][]{Megeath2012}
      {\bf (c)} {\it Herschel} map \citep[36"; ][]{Lombardi2014}; 
      {\bf (d)} ArTeMis map \citep[8"; ][]{Schuller2021};
      {\bf (e)} $N_\mathrm{{tot}}(\mathrm{H_2)}$ map (4.5"; this work).
      To facilitate their comparison, all column density maps are displayed within the same column density range.
      }
\label{fig:OMC3_NH2vsOthers}
\end{figure*}

The combination of a large dynamic range and high resolution of our $N_\mathrm{tot}(\mathrm{H_2)}$ maps show the internal cloud structure with unprecedented detail. 
As seen in Fig.\ref{fig:OMC3_NH2vsOthers} (panels c-d), the quality of our $N_\mathrm{tot}(\mathrm{H_2)}$ maps is particularly remarkable in comparison with previous single-dish studies in OMC-3. 
Our $N_\mathrm{{tot}}(\mathrm{H_2)}$ maps closely reproduce the column density peaks and large scale variations detected in continuum. 
On the other hand, most of the apparently smooth filaments, clumps, and cores reported at scales of $\sim$~0.1~pc at {\it Herschel} \citep[36" or $\sim 15000$~au,][; Fig.\ref{fig:OMC3_NH2vsOthers}c]{Lombardi2014} or even ArTeMiS \citep[8" or $\sim 3000$~au, ][; Fig.\ref{fig:OMC3_NH2vsOthers}d]{Schuller2021} resolutions resolve into sharper and smaller features when observed by ALMA (4.5" or $\sim 2000$~au; Fig.\ref{fig:OMC3_NH2vsOthers}e). 
The large dynamic range of these $N_\mathrm{{tot}}(\mathrm{H_2)}$ measurements (over two orders of magnitude) also reveals the fine structure of diffuse gas only intuited in previous column density maps. 
Similar conclusions can be drawn when comparing {\it Herschel} and $N_\mathrm{{tot}}(\mathrm{H_2)}$ maps in the other four EMERGE ALMA fields (see Appendix~\ref{appendix:DP}).
These results highlight the need of high sensitivity and high spatial resolution molecular maps for these type of studies.

\newpage 

\subsection{Caveats deriving molecular conversion factors}\label{sec:caveats}

In this paper, we introduce a new method to calibrate the observed emission of key molecular tracers into absolute gas column densities and apply it to successfully model the observed structure of five star-forming regions in Orion. This empirical approach simplifies previous treatments requiring advanced radiative transfer calculations and modeling \citep[e.g.,][]{Segal2024,Palud2025} or machine-learning techniques \citep{Gratier2021,Shimajiri2023_H2}.
Despite its potential for other studies (see also Sect.~\ref{sec:tomography}), the application of these conversions in other clouds should consider several caveats.

First, our method compares the observe line intensity of specific tracers of choice to independent estimates of the total gas column density $N\mathrm{(H_2)}$. This approach anchors our calibration to the available {\it Herschel} column density dust maps in each case.
The validity of the resulting conversion X-factors thus depend on the accuracy of these column density maps and inherits all potential biases in them (e.g., calibration of zero-extinction point due to background FIR emission). Comparisons with other datasets support the robustness of this approach in nearby clouds. Yet, its application in targets at further distances might be biased due to the intrinsically higher background contamination across the Galactic Plane \citep{Molinari2010}.

Second, and as demonstrated in Sect.~\ref{sec:Xfactors}, the integrated intensity of some of these molecular tracers is directly affected by temperature variations within clouds that need to be locally corrected to derive accurate conversion X-factors.
This is particularly relevant in the case of diffuse tracers such as HNC, more affected by external heating than N$_2$H$^+$ found deeper into the cloud (see Fig.~\ref{fig:Xfactors}).
Our technique thus requires systematic estimates of the gas kinetic temperature $T_K$ at comparable, although not necessary the same, resolution (see Sect.~\ref{sec:Xfactors_highres}). 
This might increase the observing time required in such studies to include $T_K$[HCN/HNC] \citep{Hacar2020} and/or $T_K$[NH$_3$] \citep{Friesen2017} estimates, among others, across the entire field of interest\footnote{We remark that dust temperature estimates are not valid for these corrections as the gas effective temperature $T_{dust}$ usually does not reflect the actual gas kinetic temperature, i.e. $T_K\ne T_{dust}$, unless thermally coupled at high densities $n(H_2)>10^4~\mathrm{cm^{-3}}$ \citep{Goldsmith2001}; see also \citet{Hacar2020} for a discussion.}.

Third, it is worth noticing that the above empirical calibrations aim to simplify the complex interplay between chemistry, excitation, and opacity effects affecting the different tracers considered in this analysis.
The exact calibration might then vary depending on the Galactic environment due to, for instance, local variations of their molecular 
abundance or external irradiation. 
Independent studies confirm the validity of Eq.~\ref{eq:X_N2H+} (within a factor of 2) when describing the N$_2$H$^+$ conversion factor in different environments \citep{Peretto2023,Anderson2026}.
Additional tests in other high-mass star-forming regions indicate that a local calibration of the HNC luminosity might be needed to adjust the exact slope of this conversion X-factor (Hacar et al, in prep.).

Fourth, every line-of-sight observed in a molecular cloud likely pierces throughout gas under different density and temperature conditions. 
A realistic treatment of the radiation emanating from such gas columns thus requires a complex modeling of the resulting varying excitation conditions $T_{ex}=f(n,T_K)$ along the light path using appropriate radiative transfer codes and physically motivated cloud models \citep[e.g., see][]{Tafalla2002}.
Instead, our conversion factors (Eqs.~\ref{eq:X_HNC}-\ref{eq:X_N2H+}) assume single and thermalized excitation conditions ($T_{ex}=T_K$ in Eqs.~\ref{eq:RTE}-\ref{eq:I_norm}) along each line-of-sight. 
This approach is supported by the expected proportionality between the effective column and volume density $N(\mathrm{H}_2)_{eff}\propto n(\mathrm{H}_2)^{\gamma}$ reported in recent studies \citep[][]{Gaches2025}.
Nonetheless, and despite the good agreement of our predictions at different resolutions (Sect.~\ref{sec:Xfactors_highres}), the use of a single excitation is likely responsible for the intrinsic dispersion of our data seen in Figs.~\ref{fig:Xfactors} and \ref{fig:Xfactors_highres}.

Fifth, the dynamic range of the derived column density maps is limited by the intrinsic gas regimes traced by each of the selected species, in our case HNC for the low column density gas between $N\mathrm{(H_2)}$~$\sim (2-100)\times 10^{21}$~cm$^{-2}$ and N$_2$H$^+$ for the colder and denser component at $N\mathrm{(H_2)}$~$\gtrsim 15\times 10^{21}$~cm$^{-2}$ (see Fig.~\ref{fig:Xfactors_highres}).
Previous studies noticed that the contributions of distinct molecular tracers when predicting $N\mathrm{(H_2)}$ depends on the specific column density regime considered (e.g. low, intermediate, or high column density) and gas conditions (e.g. irradiation) targeted in each case \citep{Gratier2021}.
We note however that this same approach can be extended to other diffuse (e.g., C$^{18}$O) and dense (e.g. deuterated) species if appropriate transfer functions were found, that is $N_\mathrm{{tot}}(\mathrm{H_2})=\sum_j N_j(\mathrm{H_2})\times \mathcal{W}(j)$ similar to Eq.~\ref{eq:X_total} allowing to extend this method further into a large column density range.

And sixth, these molecular line-based gas column density estimates largely improve the sensitivity and dynamic range of ALMA continuum maps of similar resolution (see also Fig.~\ref{fig:EMERGE_maps}f).
In fact, the use of such molecular proxy appears as the only viable way to overcome the intrinsic spatial filetering effects in interferometric continuum observations \citep[see][for a discussion]{Ossenkopf2016}.
In its current implementation, however, these column density estimates miss the contribution of several compact sources (i.e. YSOs) clearly detected in continuum. This is so as none of the included molecular tracers are sensitive to these objects. Future implementation of this methodology might also consider the addition of these compact sources.

\section{Multi-Tracer Column Density Synthesis}\label{sec:tomography}


Sect.~\ref{sec:NH2colmaps} demonstrates how our empirically calibrated temperature-corrected molecular conversion factors can be systematically used to derive accurate estimates of the gas column density at different resolutions. 
This novel technique allows to characterize both low column density $N_\mathrm{{HNC}}\mathrm{(H_2)}$ (Eq.~\ref{eq:X_HNC}) and high column density $N_\mathrm{{N_2H^+}}\mathrm{(H_2)}$ (Eq.~\ref{eq:X_N2H+}) gas components independently, as well as combine them into the total gas column density $N_\mathrm{{tot}}\mathrm{(H_2)}$ (Eq.~\ref{eq:X_total}) in ALMA studies.
Given the ability of this technique to reconstruct the gas column density at high spatial resolution from the combination of multiple molecular species, we dub this method as ``Multi-Tracer Column Density Synthesis'' (MT-CDS).

This MT-CDS method complements previous techniques to retrieve the total gas column density from the analysis of molecular line surveys \citep{Gratier2021,Shimajiri2023_H2,Segal2024,Palud2025}. 
Several key aspects distinct this new empirical approach from previous works.
First, MT-CDS simplifies the complex treatment of hyper-spectral machine learning techniques by the use of dedicated molecular conversion factors (Sect.~\ref{sec:Xfactors}).
Second, and related to the latter, MT-CDS specifically introduces a temperature correction that reduces the variance between regions and  allows the derivation of unique molecular conversion factors (Sect.~\ref{sec:accuracy}).
Third, our new method extends the retrieval of accurate column density measurements beyond previous estimates and into high column density regimes above $N\mathrm{(H_2)>10^{22}~cm^{-2}}$ (Sect.~\ref{sec:Xfactors_highres}).
And fourth, MT-CDS can be extrapolated to synthesize high dynamic range column density maps at interferometric ALMA resolutions of few arcsec (Sect.~\ref{sec:NH2colmaps}) providing a dramatic improve with respect to the millimeter continuum measurements at this same resolution (see also Fig.~\ref{fig:EMERGE_maps}f).

To further validate this approach, we have compared our MT-CDS column density estimates in OMC-3 with previous  {\it Spitzer} MIR observations in Fig.~\ref{fig:OMC3_NH2vsOthers}~(panels a,b). 
From the analysis of the MIR absorption seen by {\it Spitzer} (2''),
\citet{Juvela2023} derived the distribution of the dust opacity at 8~$\mu m$ ($\tau_{8\mu m}$) in this cloud.
We find a close agreement between strength and spatial distribution of these $\tau_{8\mu m}$ features and the total gas column density retrieved by our MT-CDS maps, in particular above $N_\mathrm{{tot}}\mathrm{(H_2)}\gtrsim 5\times10^{22}~\mathrm{cm^{-2}}$ (see also Appendix~\ref{appendix:Spitzer}). 
The strong similarities with these independent {\it Spitzer} estimates support the use of this MT-CDS method to accurately reproduce (synthesize) $N_\mathrm{{tot}}\mathrm{(H_2)}$ at higher spatial resolution than originally calibrated. 
Similar JWST observations could be used to test these results at increasing resolution.

Previous studies have reported the presence of narrow (0.03-0.05~pc) and highly extincted (IR-dark) gas features across the main spine of OMC-3 \citep{Juvela2023,Mannfors2025}. 
More strikingly, we find an excellent correspondence between all gas sub-structure across our $N_\mathrm{{tot}}\mathrm{(H_2)}$ maps and these IR-dark features recognized in {\it Spitzer} down to the resolution of our observations (4.5" or $\sim$2000~au).
In addition to the narrow fibers in this cloud contributing to the $N_\mathrm{{N_2H^+}}\mathrm{(H_2)}$ component \citep{SocciPaperIII}, we recognize several diffuse ``whiskers'' that emanate from regions of high-column density material towards the more diffuse gas in several parts of OMC-3 mostly seen in $N_\mathrm{{HNC}}\mathrm{(H_2)}$. These small gas features appear to be aligned with the orientation of the magnetic field reported in recent high-resolution SOFIA-HAWK+ observations \citep{Zielinski2022}. Their properties resemble the so-called ``striations'' identified at parsec-scales in the low column density gas in low-mass clouds \citep{Goldsmith2008,Palmeirim2013,Heyer2016} but with widths of few thousand au and at higher column density.

The complex gas organization detected by our $N_\mathrm{{tot}}\mathrm{(H_2)}$ maps 
highlight the fractal nature of the gas in molecular clouds \citep[e.g.,][]{Falgarone1991}.
The $N_\mathrm{{tot}}\mathrm{(H_2)}$ maps in our EMERGE ALMA Early sample shows a large variety of elongated (filamentary) features with different lengths, widths, and branches combined with other more amorphous and irregular gas shapes many of them still unresolved at 2000~au resolution.
This fine sub-structure seen by ALMA in Fig.\ref{fig:OMC3_NH2vsOthers} is hard to reconcile with a over-simplified description based on homogeneous cylinders \citep{Ostriker1964,Schneider1979}, even less with the idea of a single and ``characteristic'' width for filaments as suggested in previous {\it Herschel} studies \citep{Arzoumanian2011,Andre2014}.
Instead, the comparison between panels a-c in Fig.\ref{fig:OMC3_NH2vsOthers} reinforces the notion of a hierarchical and multi-scale nature of the ISM gas where large structures appear to fragment into smaller ones when observed at higher resolutions \citep[see][for a discussion]{Hacar2022}. 

We expect this new MT-CDS approach to open new avenues for the study of molecular clouds at high spatial resolution in the near future.
It is worth mentioning that the combination of multiple tracers make our line-based synthetic maps sensitive to a dynamic range in column density similar to those derived by {\it Herschel} in FIR continuum. 
The above $N_\mathrm{{tot}}\mathrm{(H_2)}$ maps can be used to infer the detailed mass distribution of clouds and high-dynamic range, column density Probability Distribution Functions \citep[N-PDF;][]{Kainulainen2009,Schneider2011} with ALMA \citep{Ossenkopf2016}.

The individual access to each of these $N_\mathrm{{HNC}}\mathrm{(H_2)}$ and  $N_\mathrm{{N_2H^+}}\mathrm{(H_2)}$ components can also be applied to efficiently reconstruct the true 3D structure of clouds such as OMC-3.
The combination of these and similar density selective tracers (see Sect.~\ref{sec:intro}) could potentially be employed to investigate different gas regimes within clouds: e.g., low (C$^{18}$O), intermediate (HNC), and high ($\mathrm{N_2H^+}$) volume density gas $n(\mathrm{H_2})$. 
Also, each these components can be associated to the properties of their corresponding spectra (velocity structure and dispersion, excitation, etc), in our case HNC (1-0) \citep{BonanomiPaperVI}
and N$_2$H$^+$~(1-0) \citep{SocciPaperIII}, to resolve key physical properties such as the turbulent state and gas motions (among others) of each of these components independently. Some of these topics will be explored in future papers of this series. 

\section{Summary}
This paper introduces a novel and empirically-based ``Multi-Tracer Column Density Synthesis'' (MT-CDS) technique to obtain high-dynamic range column density estimates at interferometric resolution. The main results of this work can be summarized as follows:

\begin{enumerate}

    \item We used data from the EMERGE Early ALMA Survey (Paper I) to investigate the molecular emission in 7 star-forming regions in Orion, namely OMC-1, OMC-2, OMC-3, OMC-4 South, LDN~1641N, NGC~2023, and the Flame Nebula, tracing different cloud conditions.
    There regions were homogeneously surveyed combining large-scale single-dish (IRAM-30m; 30") and interferometric (ALMA-12m array, 4.5") observations at 3~mm in continuum and molecular transitions such as N$_2$H$^+$ (1-0) and HNC (1-0).
    We complemented these observations with additional archival {\it Herschel} and {\it Spitzer} IR surveys and maps of the gas kinetic temperature $T_K$[HCN/HNC] in these regions (Sect.~\ref{sec:EMERGE}).  

    \item In agreement with previous results, the selected molecular tracers sample distinct gas regimes in our targets. N$_2$H$^+$ highlights the densest and coldest gas in our clouds at column densities $N\mathrm{(H_2)}\gtrsim 20\times 10^{21}$~cm$^{-2}$, while HNC efficiently traces the lukewarm cloud material at low and intermediate densities down to $N\mathrm{(H_2)}\sim 5\times 10^{21}$~cm$^{-2}$. 
    The integrated emission of each of these tracers is primarily driven by the total gas column density $N\mathrm{(H_2)}$ and is locally modulated by the gas kinetic temperature $T_K$ structure of our clouds. (Sect.~\ref{sec:Xfactors})
    
    \item We obtained temperature-corrected, conversion factors between the N$_2$H$^+$ and HNC integrated intensities obtained in our IRAM-30m data and the total gas column densities measure in previous {\it Herschel} observations (Sect.\ref{sec:Xfactors}). These conversions allow us to obtain individual H$_2$ column density estimates of both high $N_\mathrm{N_2H^+}\mathrm{(H_2)}$ (Eq.\ref{eq:X_N2H+}) and low $N_\mathrm{N_2H^+}\mathrm{(H_2)}$ (Eq.\ref{eq:X_HNC}) column density components, respectively. These measurements can be also combined to obtain similar estimates of the total gas column density $N_\mathrm{tot}\mathrm{(H_2)}$ (Eq.\ref{eq:X_total}) using an appropriate transfer function (Eq.~\ref{eq:X_totalW}). 
    
    \item Despite being inferred at low-resolution of single-dish data (30"), we demonstrate how these conversion factors can be transferred into higher resolutions. Comparisons with independent column density ArTeMiS maps (8") demonstrate the application and robustness of this technique using ALMA (4.5") observations (Sects.\ref{sec:Xfactors_highres}-\ref{sec:accuracy}). 
    
    \item We applied our molecular conversion factors to our ALMA+IRAM-30m data and produce independent $N_\mathrm{{N_2H^+}}\mathrm{(H_2)}$, $N_\mathrm{{HNC}}\mathrm{(H_2)}$, and $N_\mathrm{{tot}}\mathrm{(H_2)}$ column density maps at $\sim$~2000~au in five of our ALMA targets where interferometric data were available (Sect.\ref{sec:NH2maps}). 

    \item 
    We dub this empirical technique as ``Multi-Tracer Column Density Synthesis'' (MT-CDS) to capture the combination of multiple molecular species to reconstruct the total gas column density at different resolutions. Yet, this technique provides information of the gas morphology, column density, and mass load of each these components independently. This approach can also be adapted to explore other star-forming regions (Sect.\ref{sec:caveats}). 
    
    \item The enhanced sensitivity and dynamic range of our MT-CDS maps at the ALMA resolution (between $\sim 10^{21.5}-10^{23.5}$~cm$^{-2}$) reveal the fine gas organization only intuited at lower resolutions. 
    As confirmation of these results, we find remarkable similarities between our $N_\mathrm{{tot}}\mathrm{(H_2)}$ maps and most of the extinction features seen by {\it Spitzer} at MIR wavelengths.
    These new column density maps at high spatial resolution measurements highlight the hierarchical and multi-scale nature of the ISM with unprecedented detail (Sect.~\ref{sec:tomography}).

\end{enumerate}

\begin{acknowledgements}
      This project has received funding from the European Research Council (ERC) under the European Union’s Horizon 2020 research and innovation programme (Grant agreement No. 851435).
      This paper makes use of the following ALMA data: ADS/JAO.ALMA\#2019.1.00641.S. ALMA is a partnership of ESO (representing its member states), NSF (USA) and NINS (Japan), together with NRC (Canada), MOST and ASIAA (Taiwan), and KASI (Republic of Korea), in cooperation with the Republic of Chile. The Joint ALMA Observatory is operated by ESO, AUI/NRAO and NAOJ.
      This work is based on IRAM-30m telescope observations carried out under project numbers 032-13, 120-20, 060-22, and 133-22. IRAM is supported by INSU/CNRS (France), MPG (Germany), and IGN (Spain). 
      This research has made use of the SIMBAD database, operated at CDS, Strasbourg, France.
      This research has made use of NASA’s Astrophysics Data System.
      This work made use of Astropy:\footnote{http://www.astropy.org} a community-developed core Python package and an ecosystem of tools and resources for astronomy \citep{astropy:2013, astropy:2018, astropy:2022}. This work made intensive use of the python packages {\it Matplotlib} \citep{matplotlib:2007} and {\it Numpy} \citep{Numpy:2020}.
\end{acknowledgements}
\bibliographystyle{aa}
\bibliography{filaments.bib}

\begin{appendix}
    
\section{Data products}\label{appendix:DP}

\begin{figure*}[t]
\centering
\includegraphics[width=1.\textwidth]{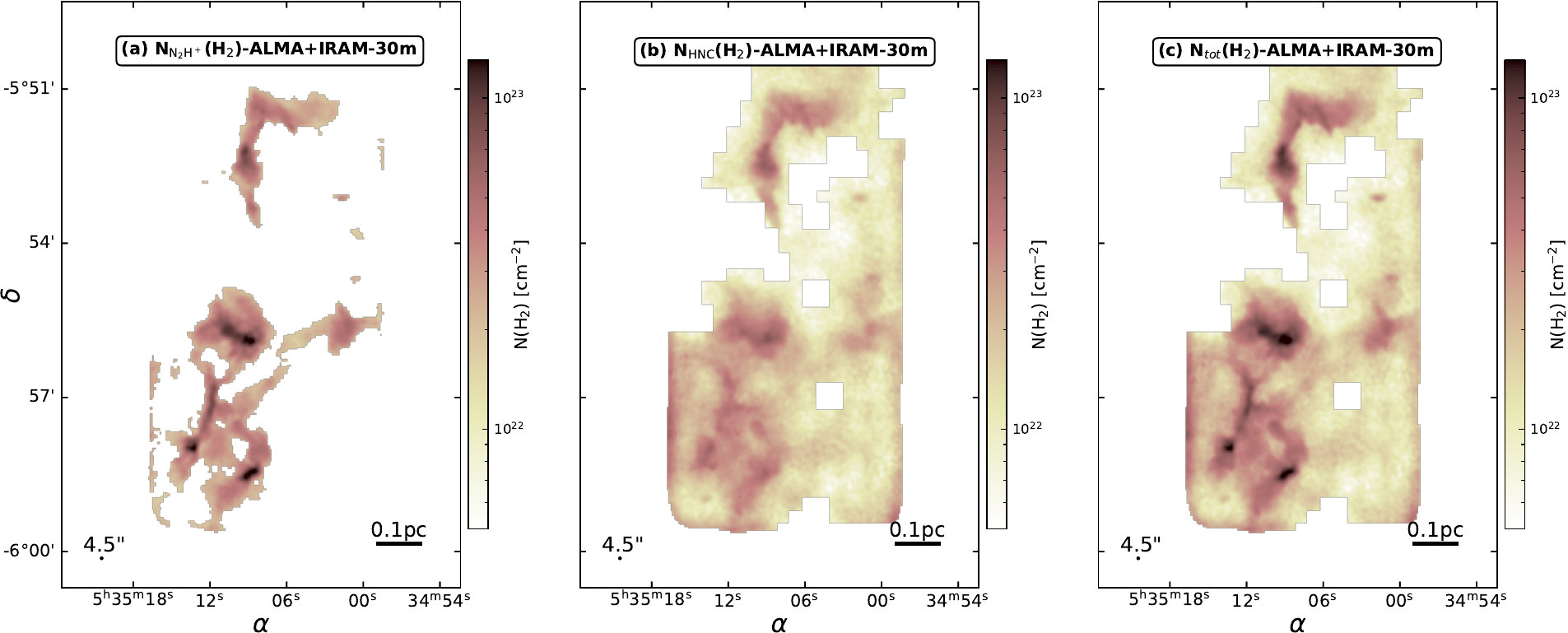}
      \caption{High-resolution (4.5") column density estimates in OMC-4~South. Panels and symbols are similar to Fig.\ref{fig:OMC3_NH2}.
      }
\label{fig:DP_OMC4_NH2}
\end{figure*}

\begin{figure*}[t]
\centering
\includegraphics[width=1.\textwidth]{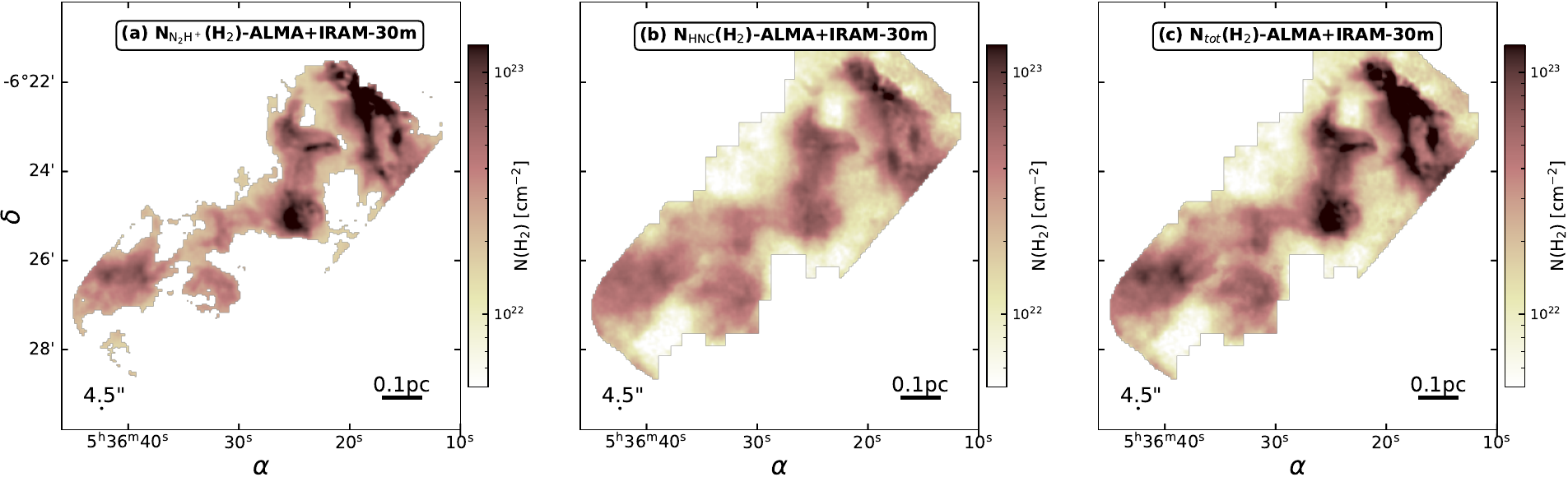}
      \caption{High-resolution (4.5") column density estimates in LDN~1641N. Panels and symbols are similar to Fig.\ref{fig:OMC3_NH2}.
      }
\label{fig:DP_LDN1641N_NH2}
\end{figure*}

\begin{figure*}[t]
\centering
\includegraphics[width=1.\textwidth]{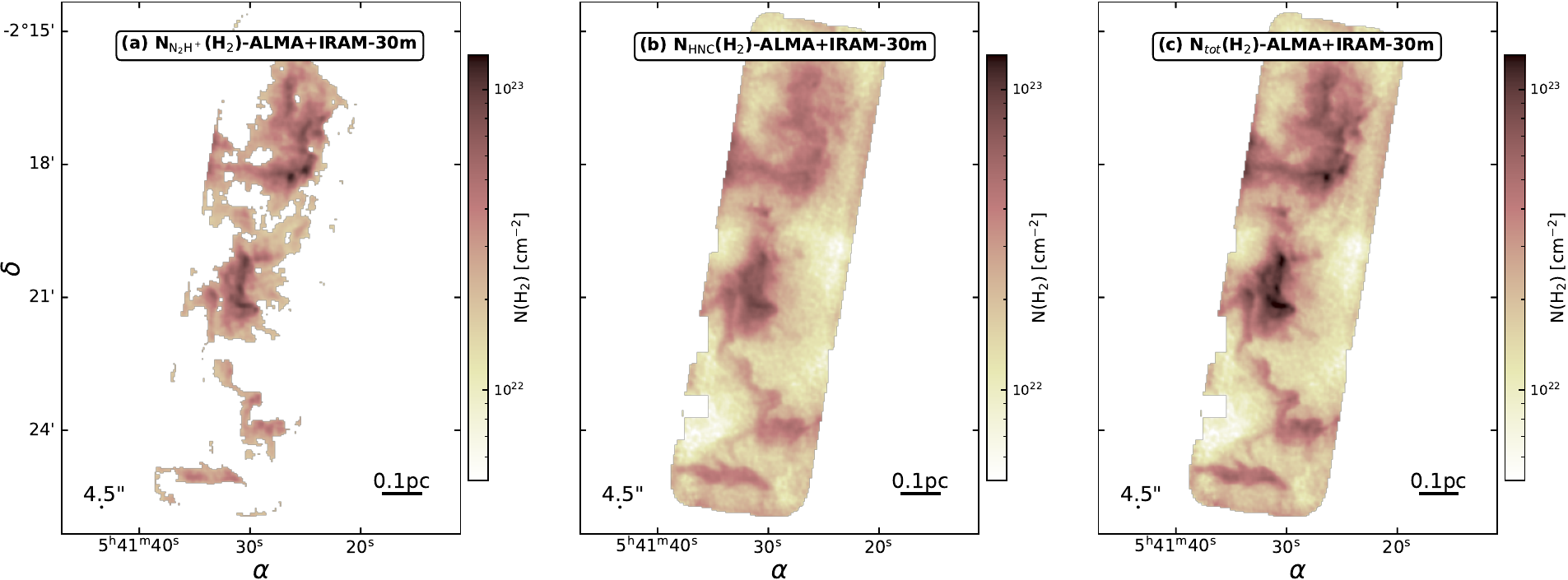}
      \caption{High-resolution (4.5") column density estimates in NGC~2023. Panels and symbols are similar to Fig.\ref{fig:OMC3_NH2}.
      }
\label{fig:DP_NGC2023_NH2}
\end{figure*}

\begin{figure*}[t]
\centering
\includegraphics[width=1.\textwidth]{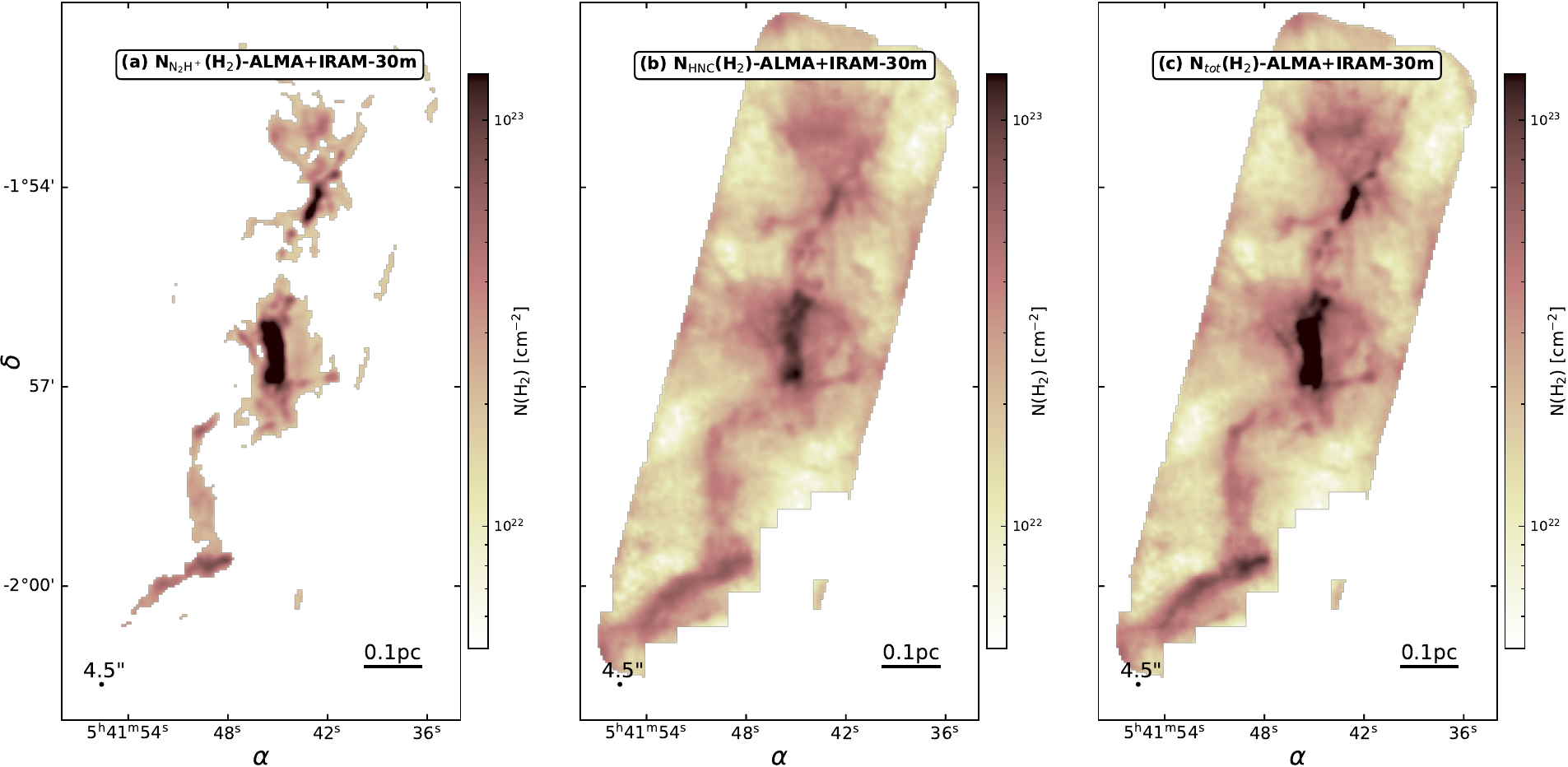}
      \caption{High-resolution (4.5") column density estimates in Flame Nebula. Panels and symbols are similar to Fig.\ref{fig:OMC3_NH2}.
      }
\label{fig:DP_FlameNeb_NH2}
\end{figure*}

Similar to Fig.~\ref{fig:OMC3_NH2}, Figs~\ref{fig:DP_OMC4_NH2}-\ref{fig:DP_FlameNeb_NH2} show the resulting $N_\mathrm{N_2H^+}\mathrm{(H_2)}$ (left panel), $N_\mathrm{HNC}\mathrm{(H_2)}$ (central panel), and $N_\mathrm{tot}\mathrm{(H_2)}$ (right panel) estimates at 4.5" resolution in all other regions (OMC-4 South, LDN~1641N, NGC~2023, and Flame Nebula) of our survey in which all HNC and N$_2$H$^+$ ALMA+IRAM30m maps, together with $T_K[\mathrm{HCN/HNC}]$ estimates from IRAM-30m, were available. Note that some diffuse positions traced by HNC are missing in these images where no $T_K[\mathrm{HCN/HNC}]$ data were available at these locations (see temperature maps in Paper I).


\section{Comparison with Spitzer MIR opacity}\label{appendix:Spitzer}

\begin{figure*}[ht]
\centering
\includegraphics[width=1.\textwidth]{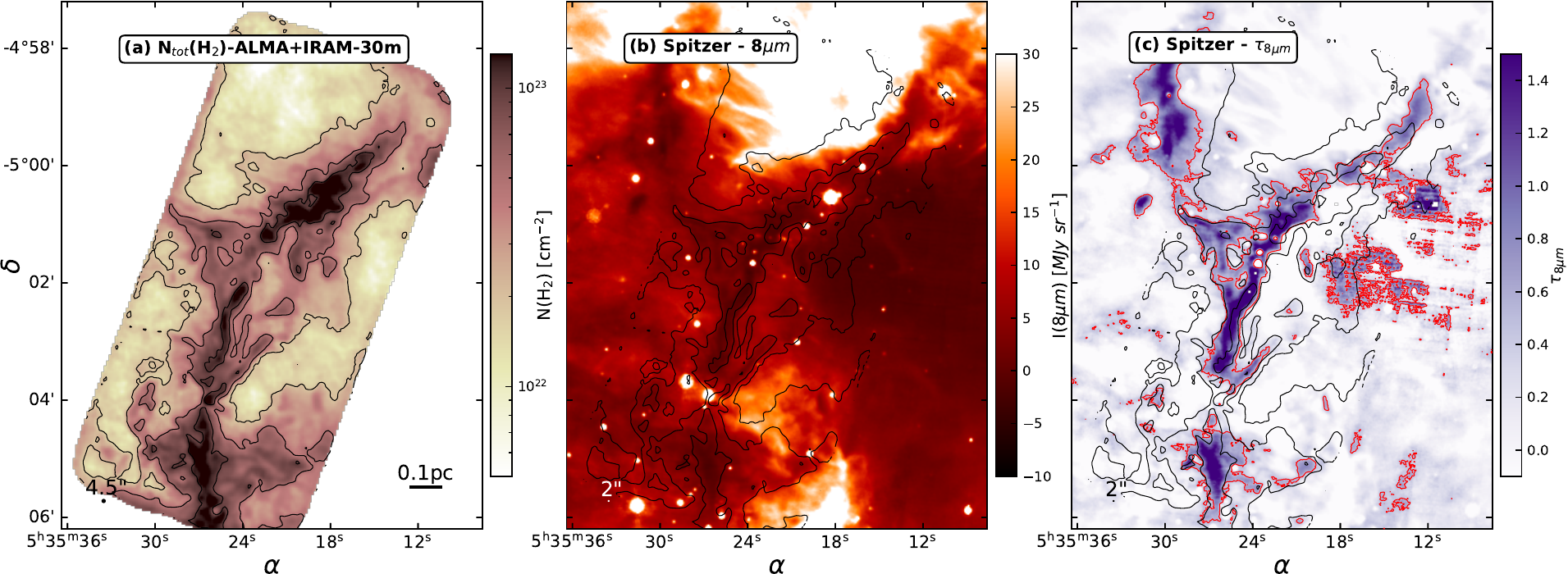}
      \caption{
      From left to right and from top to bottom: 
      {\bf (a)} MT-CDS $N_\mathrm{tot}(\mathrm{H_2)}$ total column density maps obtained from our ALMA+IRAM-30m observations (4.5"), 
      {\bf (b)} {\it Spitzer} 8$\mu m$ emission and
      {\bf (c)} $\tau_{8\mu m}$ opacity \citep{Juvela2023}. 
      In all panels we overplot the contours of our MT-CDS maps at $N_\mathrm{{tot}}(\mathrm{H_2)}=[2,5,10]\times10^{22}~\mathrm{cm^{-2}}$, respectively (black lines).
      Panel c also includes the contour for which $\tau_{8\mu m}\ge0.5$ (red line).
      }
\label{fig:OMC3_Spitzer}
\end{figure*}

In Figure~\ref{fig:OMC3_Spitzer} compares the column density estimates retrieved by our our MT-CDS technique in OMC-3 (Fig.~\ref{fig:OMC3_Spitzer}a; 4.5'') with the 8~$\mu m$ emission $I(8\mu m)$ (Fig.~\ref{fig:OMC3_Spitzer}b) and opacity $\tau_{8\mu m}$ (Fig.~\ref{fig:OMC3_Spitzer}c) reported by \citet{Juvela2023} using {\it Spitzer} observations (2'').
We notice the close correspondence between the our $N_\mathrm{{tot}}(\mathrm{H_2)}$ total gas column density estimates at high spatial resolution and the highly extincted regions in the MIR ($\tau_{8\mu m}\ge 0.5$; red contour in panel c), particularly above $N_\mathrm{{tot}}(\mathrm{H_2)}\ge 5\times10^{22}~\mathrm{cm^{-2}}$ (second black contour in all images). Differences between these maps can be explained by the non-uniform illumination of this region as seen by {\it Spitzer} \citep[see][for a discussion]{Juvela2023}.

\end{appendix}

\end{document}